\documentclass[journal]{IEEEtran}

\makeatletter
\def\ps@headings{%
	\def\@oddhead{\mbox{}\scriptsize\rightmark \hfil \thepage}%
	\def\@evenhead{\scriptsize\thepage \hfil \leftmark\mbox{}}%
	\def\@oddfoot{}%
	\def\@evenfoot{}}
\makeatother 

\makeatletter
\def\endthebibliography{%
	\def\@noitemerr{\@latex@warning{Empty `thebibliography' environment}}%
	\endlist
}
\makeatother

\usepackage{graphicx}
\usepackage{makecell} 
\usepackage{amsmath, amsfonts,epsfig, multirow, floatflt}
\usepackage{amssymb}
\usepackage{subfigure}
\usepackage{cite}
\usepackage{algorithm}
\usepackage[normalem]{ulem}

\usepackage{placeins}
\usepackage{algorithmicx}
\usepackage{algpseudocode}
\usepackage{hyperref}
\usepackage{enumitem}
\usepackage{diagbox}
\usepackage{mathrsfs}
\usepackage{url}
\usepackage{color}

\usepackage{caption}
\usepackage{hhline}
\usepackage{array}
\usepackage{booktabs}
\usepackage{amsthm}
\usepackage{multirow}

\usepackage{bm}
\usepackage{mathtools}
\usepackage{siunitx}
\usepackage{booktabs}
\usepackage{setspace}
\usepackage[table,xcdraw]{xcolor}
\definecolor{ochre}{RGB}{184,134,11} 
\definecolor{navy}{RGB}{0,0,128} 
\usepackage{comment}

\allowdisplaybreaks[4]
\graphicspath{{figure/}}
\begin{document}
	
	\title{Multi-Agent Reinforcement Learning for V2X Resource Allocation: Disentangling MARL Challenges Through Benchmarking} \author{Siyuan~Wang$^{1}$, Lei~Lei$^{1}$,~\IEEEmembership{Senior~Member,~IEEE}, Pranav~Maheshwari$^{1}$, Sam~Bellefeuille$^{1}$, Kan~Zheng$^{2}$ {\it Fellow, IEEE }

%\thanks {An earlier version of this work appeared at the IEEE International Conference on Communications (ICC) 2025 \cite{wang2025multiagent}. This version extends the prior work by enlarging and systematizing the training and testing datasets with open-source dataset and code to enable representative and reproducible evaluations, expanding the scalability analysis to more agents, and providing deeper experimental analysis, an expanded literature review, and key insights into future work.}

\thanks {This work was supported by the Natural Sciences and
Engineering Research Council of Canada (NSERC) under Discovery Grant.}

\thanks{$^{1}$S. Wang, L. Lei, P. Maheshwari, and S. Bellefeuille are with the College of Engineering, University of Guelph, Guelph, Ontario, N1G 2W1, Canada (e-mail: leil@uoguelph.ca).}

\thanks{$^{2}$K. Zheng is with the College of Electrical Engineering and Computer Sciences, Ningbo University, Ningbo, 315211, China.}
}

%\thanks{Copyright (c) 2026 IEEE. Personal use of this material is permitted. However, permission to use this material for any other purposes must be obtained from the IEEE by sending a request to pubs-permissions@ieee.org.}

\maketitle

\begin{abstract}
Multi-agent reinforcement learning (MARL) has emerged as a promising approach to radio resource allocation (RRA) in cellular vehicle-to-everything (C-V2X) networks. However, MARL challenges such as non-stationarity, coordination difficulty, large joint action spaces, partial observability, and limited robustness and generalization are often intertwined, making it difficult to identify the key performance bottlenecks and determine which MARL paradigms address them effectively. We introduce a challenge-disentangling benchmarking framework that formulates C-V2X RRA as a hierarchy of multi-agent interference games and constructs learning tasks that progressively introduce MARL challenges, enabling their incremental performance effects to be quantified. Using vehicular topology datasets generated with the Simulation of Urban MObility (SUMO) simulator, we systematically evaluate eight representative MARL algorithms spanning value-based and actor-critic methods under Independent Learning (IL) and Centralized Training with Decentralized Execution (CTDE). Three main insights emerge. First, robustness across diverse vehicular topologies is a major performance bottleneck at larger network scales, reducing the average normalized return by up to 35 percentage points in the 16-agent setting. Second, actor-critic methods increasingly outperform value-based methods at larger network scales, particularly in more complex tasks, with an average performance advantage of up to 36.3 percentage points. Third, Independent Proximal Policy Optimization (IPPO) provides a strong and scalable baseline, achieving performance close to its CTDE counterpart while avoiding the scalability burden of centralized training. These findings identify future research directions for MARL-based C-V2X RRA, while the systematic evaluation framework and publicly available code and datasets provide a reproducible foundation for developing and evaluating new algorithms.

\end{abstract}

\begin{IEEEkeywords}
Deep Reinforcement Learning; V2X; Radio Resource Allocation; MARL
\end{IEEEkeywords}

\section{Introduction}

Cellular Vehicle-to-Everything (C-V2X) is a key technology
for intelligent transportation systems, supporting
vehicle-to-vehicle (V2V) and vehicle-to-infrastructure
(V2I) communications. Radio Resource Allocation (RRA),
which coordinates limited communication resources such
as spectrum and transmit power among vehicles and
infrastructure nodes, plays a critical role in C-V2X
network performance \cite{liu2021recent}. The dynamic
and uncertain nature of vehicular environments has
motivated the use of Deep Reinforcement Learning (DRL)
for RRA, enabling resource allocation policies to be
learned through interactions with the environment
\cite{lei2019multiuser,ye2019deep}.

RRA in C-V2X networks is inherently a multi-agent
decision-making problem, as efficient resource sharing
depends on the coordinated actions of multiple vehicles
and devices. Thus, Multi-Agent Reinforcement Learning
(MARL) has received increasing attention as a framework
for learning distributed RRA policies \cite{marl-book}.

Recent research has demonstrated the potential of MARL
for C-V2X RRA, leading to the development of increasingly
sophisticated algorithms
\cite{ye2019deep,liang2019spectrum,
Xiang2021MARLDecentralizedSA,Xiang2023EmergentComm,
Ding2022AMARL,zhang2022mean,
ji2025graphneuralnetworksdeep,Yuan2021MetaRLV2X,
lei2023multitimescale}. Although a wide range of
value-based, actor-critic, Independent Learning (IL),
and Centralized Training with Decentralized Execution
(CTDE) approaches have been applied to C-V2X RRA,
systematic comparisons among these MARL paradigms
remain limited. Comprehensive MARL benchmarks have
been developed across standard cooperative multi-agent
environments \cite{papoudakis2021benchmarking}, but
how their findings apply to C-V2X RRA remains unclear
due to its domain-specific interference and
resource-sharing structure.

More fundamentally, MARL faces several well-known
challenges \cite{oroojlooy2023review} that often coexist
and interact in C-V2X RRA. Simultaneous learning by
multiple agents introduces non-stationarity, while
interference-coupled decisions create coordination
difficulty. Additional environmental features introduce
further challenges: multiple time steps give rise to
sequential decision making, fast fading introduces
environmental stochasticity, increasing network scale
enlarges the joint action space, varying vehicular
topologies challenge policy robustness and
generalization, and local observations introduce
partial observability. These challenges may interact,
such that additional challenges can intensify those
already present. Conventional comparisons typically
evaluate algorithms in environments where multiple
challenges coexist, making it difficult to determine
their relative impacts on learning performance and
how effectively different MARL approaches address them.

Motivated by these observations, this paper seeks
to answer the following research questions:
\begin{itemize}
    \item What is the relative impact of different
    MARL challenges on C-V2X RRA performance?
    \item How do different classes of MARL algorithms
    compare in addressing these challenges?
\end{itemize}

To address these questions, the main contributions
of this work are summarized as follows.

\begin{itemize}
    \item We formulate C-V2X RRA as a hierarchy of
    multi-agent interference games that progressively
    incorporates additional environmental features,
    including multiple time steps, fast fading,
    increasing network scale, varying vehicular
    topologies, and local observations.
    Based on this framework, we develop a suite of
    benchmark learning tasks to systematically
    investigate the incremental performance effects
    associated with the corresponding MARL challenges through controlled
    comparisons. We further construct Simulation of
    Urban MObility (SUMO)-based training and evaluation
    datasets spanning diverse vehicle densities and
    interference conditions.

    \item Using the proposed benchmark, we
    systematically evaluate eight representative
    MARL algorithms spanning value-based and
    actor-critic methods under IL and CTDE.
    Cross-task and cross-algorithm comparisons reveal
    the strengths and limitations of different MARL
    paradigms in addressing the considered challenges.
    The results identify learning robust policies
    across diverse vehicular topologies as a major
    performance bottleneck at larger network scales
    and establish Independent Proximal Policy
    Optimization (IPPO) as a strong and scalable
    baseline for C-V2X RRA. We release the complete
    benchmark suite, including datasets and
    code\footnote{\url{https://github.com/Deepinlab2023/V2X-MARL-Bench}},
    providing a systematic and reproducible foundation
    for evaluating and advancing MARL algorithms in
    vehicular networks.
\end{itemize}

An earlier version of this work appeared at the
IEEE International Conference on Communications
(ICC) 2025~\mbox{\cite{wang2025multiagent}}.
This journal paper substantially extends the
conference version from an algorithm comparison
to a comprehensive challenge-oriented benchmarking
framework, introducing learning tasks designed
to investigate additional MARL challenges,
large-scale SUMO-generated datasets based on
standardized Third Generation Partnership Project
(3GPP) and European Telecommunications Standards
Institute (ETSI) evaluation scenarios
~\mbox{\cite{3gpp_release14,ETSITR103766}},
expanded analysis of key MARL challenges, and
a publicly released benchmark suite with datasets
and source code.

The remainder of this paper is organized as follows.
Section~II reviews related work, and Section~III
presents the C-V2X communication system model.
Section~IV formulates the RRA problem as a hierarchy
of multi-agent interference games and identifies
the associated MARL challenges. Section~V introduces
the eight MARL algorithms under evaluation and
analyzes their strengths and limitations.
Section~VI describes the experimental methodology,
followed by the results and discussion in
Section~VII. Finally, Section~VIII presents
the conclusions and future research directions.

\section{Related Works}
MARL has been extensively investigated for RRA in C-V2X communications. This section reviews representative studies from two perspectives: the employed DRL algorithms and the C-V2X evaluation environments. Table~\ref{tab:literature_review} summarizes their main characteristics.

\begin{table*}[t]
\centering
\caption{Literature Review of MARL Applications in C-V2X Environments}
\label{tab:literature_review}
% \resizebox{\textwidth}{!}{%
\scriptsize
\begin{tabular}{ccccccc}
\toprule
\textbf{Reference} & \textbf{Scenario} & \textbf{\# V2V} & \textbf{\# V2I} & \textbf{Speed} &
\textbf{Payload} & \textbf{DRL Algorithm} \\
\midrule
% ============================================
% Ordered by citation appearance in Related Works
% Section II.A (DRL Algorithms) first
% ============================================
% [2] - H. Ye and G. Y. Li
\cite{ye2019deep} & Urban & 60-300 & - & 36 km/h &
\begin{tabular}[c]{@{}l@{}}Train: -\\Test: -\end{tabular} &
DQN \\[0.3em]
\hline
% [4] - L. Liang et al.
\cite{liang2019spectrum} & Urban & 4 & 4 & 36 km/h &
\begin{tabular}[c]{@{}l@{}}Train: $2\times1060$B\\Test: $(1,\dots,6)\times1060$B\end{tabular} &
DQN (fingerprint) \\[0.3em]
\hline
% [8] - P. Xiang et al.
\cite{Xiang2021MARLDecentralizedSA} & Urban & 4/8 & 4/8 & 36--54 km/h &
\begin{tabular}[c]{@{}l@{}}Train: $6\times1060$B\\Test: $(1,\dots,6)\times1060$B\end{tabular} &
Hysteretic D3RQN \\[0.3em]
\hline
% [9] - P. Xiang et al. (emergent communication)
\cite{Xiang2023EmergentComm} & Urban & 4 & 4 & 36/54 km/h &
\begin{tabular}[c]{@{}l@{}}Train: $(1,\dots,6)\times210$B\\Test: $(1,\dots,6)\times210$B\end{tabular} &
DQN (emergent comm.) \\[0.3em]
\hline
% [10] - Y. Ding et al.
\cite{Ding2022AMARL} & Urban & 4 & 4 & 36--54 km/h &
\begin{tabular}[c]{@{}l@{}}Train: $2\times1060$B\\Test: $(1,\dots,6)\times1060$B\end{tabular} &
DQN (attention) \\[0.3em]
\hline
% [11] - H. Zhang et al.
\cite{zhang2022mean} & Urban & 8-32 & - & 36--144 km/h &
\begin{tabular}[c]{@{}l@{}}Train: -\\Test: $2\times1060$B\end{tabular} &
DQN (mean-field) \\[0.3em]
\hline
% [12] - J. Tian et al.
\cite{10118603} & Urban & 4 & 4 & 10--60 km/h &
\begin{tabular}[c]{@{}l@{}}Train: $n\times1060$B\\Test:  $n\times1060$B\end{tabular} &
DDQN \\[0.3em]
\hline
% [13] - J. Gui et al.
\cite{10436060} & Urban & 5-25 & 3-12 & - &
\begin{tabular}[c]{@{}l@{}}Train: -\\Test: -\end{tabular} &
DDQN (federated) \\[0.3em]
\hline
% [14] - M. Ji et al.
\cite{ji2025graphneuralnetworksdeep} & Urban & 60-300 & - & - &
\begin{tabular}[c]{@{}l@{}}Train: -\\Test: -\end{tabular} &
DDQN (GNN) \\[0.3em]
\hline
% [18] - M. Jamal et al.
\cite{jamal2024hybrid} & Urban & 4 & 4 & 36 km/h &
\begin{tabular}[c]{@{}l@{}}Train: $(1,\dots,6)\times1060$B\\Test: $(1,\dots,6)\times1060$B\end{tabular} &
QMIX \\[0.3em]
\hline
% [19] - K. Xu et al.
\cite{xu2024federated} & Urban & 4-24 & 4-8 & 36--54 km/h &
\begin{tabular}[c]{@{}l@{}}Train: $2\times1060$B\\Test: $(1,\dots,6)\times1060$B\end{tabular} &
policy gradient (federated) \\[0.3em]
\hline
% [20] - Z. Shao et al.
\cite{shao2024semantic} & Urban & 0 & 5 & 36 km/h &
\begin{tabular}[c]{@{}l@{}}Train: - \\Test: - \end{tabular} &
MAPPO \\[0.3em]
\hline
% [21] - Y. Xu et al.
\cite{xu2023deep} & Highway (platooning) & - & 25 & 10--90 km/h &
\begin{tabular}[c]{@{}l@{}}Train: 300B \\Test: 300B \end{tabular} &
MAPPO \\[0.3em]
\hline
% [22] - Y. Song et al.
\cite{10729847} & Highway & 48 & 3 & - &
\begin{tabular}[c]{@{}l@{}}Train: 300--12000B \\Test: 300--12000B\end{tabular} &
IPPO \\[0.3em]
\hline
% [23] - M. Parvini et al.
\cite{parvini2023aoi} & Urban (platooning) & 15-45 & 5-7 & 36--54 km/h &
\begin{tabular}[c]{@{}l@{}}Train: 4000B\\Test: 4000B\end{tabular} &
MADDPG \\[0.3em]
\hline
% [26] - W. Zhang et al.
\cite{zhang2025semanticawareresourcemanagementcv2x} & Urban (platooning) & 16 & 4 & 36 km/h &
\begin{tabular}[c]{@{}l@{}}Train: -\\Test: -\end{tabular} &
MADDPG \\[0.3em]
\hline
% [27] - Y. Yuan et al.
\cite{Yuan2021MetaRLV2X} & Urban / Highway & 4-20 & 4 & 50-108 km/h &
\begin{tabular}[c]{@{}l@{}}Train: $1\times1060$B\\Test: $(1,\dots,6)\times1060$B\end{tabular} &
DQN + DDPG \\[0.3em]
\hline
\bottomrule
\end{tabular}%
% }
\\[0.4em]
\footnotesize
``-''indicates that the parameter value is not specified in the experiments. 
\end{table*}

\subsection{DRL Algorithms}
DRL algorithms can be broadly categorized as value-based, policy gradient, and actor-critic methods \cite{lei2020deep}, while multi-agent implementations commonly follow IL or CTDE frameworks.

Value-based IL has been widely adopted for C-V2X RRA \cite{ye2019deep,liang2019spectrum,Xiang2021MARLDecentralizedSA,Xiang2023EmergentComm,Ding2022AMARL,zhang2022mean,10118603,10436060,ji2025graphneuralnetworksdeep}. Building on Deep Q-Network (DQN)-based resource allocation \cite{ye2019deep}, subsequent studies have incorporated fingerprints and hysteretic learning to mitigate non-stationarity and coordination difficulty \cite{liang2019spectrum,Xiang2021MARLDecentralizedSA}, emergent communication, attention mechanisms, and graph neural networks (GNNs) to enhance inter-agent information exchange and state representation \cite{Xiang2023EmergentComm,Ding2022AMARL,ji2025graphneuralnetworksdeep}, and mean-field approximations to improve scalability \cite{zhang2022mean}. Double DQN (DDQN)-based approaches have also been investigated \cite{10118603,10436060}, while value-based CTDE methods such as QMIX have received comparatively less attention \cite{jamal2024hybrid}.

Policy gradient and actor-critic methods have increasingly been explored as alternatives to value-based approaches. Representative studies include IL and CTDE variants of Proximal Policy Optimization (PPO) \cite{shao2024semantic,xu2023deep,10729847}, as well as deep deterministic policy gradient (DDPG) and multi-agent DDPG (MADDPG)-based approaches \cite{parvini2023aoi,zhang2025semanticawareresourcemanagementcv2x}. Meta-RL and federated RL have further been investigated to improve adaptation and distributed learning \cite{Yuan2021MetaRLV2X,xu2024federated}. 

\subsection{C-V2X Evaluation Environments}
Most existing studies adopt road and channel models derived from 3GPP TR 36.885 \cite{3gpp_release14}, with urban scenarios being more extensively studied \cite{xu2024federated,10436060,zhang2025semanticawareresourcemanagementcv2x,jamal2024hybrid,Ding2022AMARL,Xiang2021MARLDecentralizedSA,10118603,ye2019deep,Yuan2021MetaRLV2X,zhang2022mean,ji2025graphneuralnetworksdeep} than highway scenarios \cite{10729847,Yuan2021MetaRLV2X}. A common RRA configuration assigns each V2I link a dedicated uplink subchannel that can be reused by V2V links \cite{ye2019deep,liang2019spectrum}, with four V2I and four V2V links being a frequently adopted setting \cite{liang2019spectrum,Xiang2021MARLDecentralizedSA,jamal2024hybrid,Ding2022AMARL,10118603}. Although some studies consider substantially larger networks \cite{ye2019deep,ji2025graphneuralnetworksdeep}, they employ asynchronous or batched agent decision making to manage the increased scale.

While 3GPP TR 36.885 \cite{3gpp_release14} and ETSI TR 103 766 \cite{ETSITR103766} specify representative vehicle speeds and densities for reproducible evaluations, many existing studies adopt vehicle speeds around 36--54 km/h \cite{ye2019deep,liang2019spectrum,Xiang2021MARLDecentralizedSA,Xiang2023EmergentComm,Ding2022AMARL,jamal2024hybrid,xu2024federated,shao2024semantic,parvini2023aoi,zhang2025semanticawareresourcemanagementcv2x}, and vehicle density is often unspecified. This omission is particularly relevant to RRA because vehicle density directly affects network topology and interference conditions. Payload configurations differ across studies, although multiples of 1,060 bytes are commonly adopted \cite{xu2024federated,jamal2024hybrid,Ding2022AMARL,Xiang2021MARLDecentralizedSA,10118603,Yuan2021MetaRLV2X,zhang2022mean}.

\subsection{Summary and Research Gap}
The diversity of DRL algorithms and experimental settings makes cross-study performance comparisons difficult. Moreover, individual studies generally evaluate a particular algorithm or a small set of baselines within a given environment, where multiple sources of learning difficulty coexist. Consequently, existing evaluations provide limited insight into the relative impact of individual MARL challenges and how representative MARL paradigms respond to them.

\section{C-V2X Communication System}

  We consider a typical C-V2X network, where V2V links coexist with V2I links \cite{3gpp_release14}. A V2I link connects a vehicle to the base station (BS) and can be used for high-throughput services. A V2V link connects a pair of neighboring vehicles for periodic transmission of cooperative awareness messages (CAM) to support advanced driving services. \par 

   In order to enhance spectrum utilization, we consider that $L$ V2V links share the uplink resources orthogonally allocated to $M$ V2I links. Without loss of generality, we assume that every V2I link $m\in \mathcal{M}=\{0,\ldots, M-1\}$ is pre-assigned subchannel $m$ with constant transmit power $P^{\rm I}_{m}$ \cite{liang2019spectrum}. Moreover, one or more V2V links $i\in \mathcal{L}=\{0,\ldots, L-1\}$ can reuse the subchannels of the V2I links for CAM transmission.  
    
 The time is discretized into equal-length control intervals, indexed by $k \in \mathcal{K} = \{0,1,\ldots\}$. At the beginning of each control interval $k$ (i.e., time $kT$), the transmitting vehicle of each V2V link $i$ (i.e., V2V transmitter $i$) samples its driving status to form the CAM and buffers it in a queue before transmitting the data to the corresponding receiving vehicle (i.e., V2V receiver $i$).\par 
    
Each control interval $k$ is further divided into $T$ communication intervals indexed by $t \in \mathcal{T}=\{0,1,\ldots, T-1\}$ on a faster timescale. Each communication interval has a length of $\Delta t$ $\rm ms$ corresponding to the subframe duration in C-V2X communications. We consider dynamic scheduling \cite{lei2023multitimescale}, where the V2V transmitters make RRA decisions at time $kT+t, k \in \mathcal{K}, t \in \mathcal{T}$, and the corresponding communication interval is represented as $(k,t)$. In the rest of the paper, we will use $x_{(k,t)}:=x((kT+t)\Delta t)$ to represent any variable $x$ at the beginning of communication interval $(k,t)$. \par

%Temporal integrity is maintained by $(k,T)=(k+1,0)$. 

%we will use $x_{k}:=x(kT)$ to represent any variable $x$ at control interval $k$ and
    
We use the binary allocation indicator $\theta_{i,m,(k,t)}\in\{0,1\}$ to indicate whether V2V link $i$ occupies subchannel $m$ at communication interval $(k,t)$ or not. Moreover, we consider that each V2V link $i$ occupies at most one subchannel, i.e., $\sum_{m=0}^{M-1} \theta_{i,m,(k,t)} \leq 1$.\par
    % Consider that a dedicated chunk of subchannels $m\in\{1,2,\dots, M\}$ is allocated to the platoon. 
 %  In each communication interval $(k,t)$, the $i$-th V2V transmitter transmits the data in its queue according to the local RRA decisions. 
    
    \subsubsection{Channel Gain}
    
    The instantaneous channel gain of V2V link $i$ over subchannel $m$ (occupied by V2I link $m$) at communication interval $(k,t)$ is denoted by $G_{i,m,(k,t)}$. $G_{i,m,(k,t)}$ remains constant within a communication interval, and can be written as
    \begin{equation}
    G_{i,m,(k,t)}=\alpha_{i,(k,t)}h_{i,m,(k,t)},
    \end{equation} 
   \noindent where $\alpha_{i,(k,t)}$ captures the large-scale fading effects including path loss and shadowing, which are assumed to be frequency independent. $h_{i,m,(k,t)}$ corresponds to small-scale fading, which can be regarded as identically distributed random variables with unit mean on all subchannels. We consider that $G_{i,m,(k,t)}$ is a Markov process \cite{lei2015delay}, where the probability of $G_{i,m,(k,t)}$ taking a specific value only depends on the value of $G_{i,m,(k,t-1)}$.\par 
    
    Similarly, let $G_{m,(k,t)}$ denote the channel gain of the V2I link $m$; $G_{i,B,m,(k,t)}$ denote the interference channel gain from V2V link $i$ transmitter to V2I link $m$ receiver; $G_{B,i,m,(k,t)}$ represent the interference channel gain from V2I link $m$ transmitter to V2V link $i$ receiver; and $G_{j,i,m,(k,t)}$ be the interference channel gain from the V2V link $j$ transmitter to the V2V link $i$ receiver over the subchannel $m$.\par
    
    \subsubsection{Signal-to-Interference-plus-Noise Ratio (SINR)}
    % We assume that there is no interference and that the V2V link does not reuse subchannels. Therefore, the higher the transmission power $p^m_{i,(k,t)}$, the better the channel state. 
    The SINR $\gamma_{m,(k,t)}$ and $\gamma_{i,m,(k,t)}$ of V2I link $m$ and V2V link $i$ on subchannel $m$ at communication interval $(k,t)$ are derived by 
    \begin{align}\label{SINRV2I}
    \gamma_{m,(k,t)} = \frac{P^{\rm I}_{m} G_{m,(k,t)}}{\sigma^2 + \sum\limits_{i\in\mathcal{L}}\theta_{i,m,(k,t)} P^{\rm V}_{i,m,(k,t)} G_{i,B,m,(k,t)}},
    \end{align}
    and
    \begin{align}\label{SINRV2V}
    \gamma_{i,m,(k,t)}= \frac{P^{\rm V}_{i,m,(k,t)} G_{i,m,(k,t)}}{\sigma^2 + I_{i,m,(k,t)}},
    \end{align}
    respectively, \noindent where $P^{\rm V}_{i,m,(k,t)}$ is the transmit power of V2V link $i$ over the subchannel $m$ at communication interval $(k,t)$. $\sigma^2$ is the power of channel noise which satisfies the independent Gaussian distribution with a zero mean value. $I_{i,m,(k,t)}$ is the total interference power received by V2V link $i$ over subchannel $m$, where
    \begin{align}\label{IV2V}
    & I_{i,m,(k,t)}=\IEEEnonumber \\
    & P^{\rm I}_{m} G_{B,i,m,(k,t)}+ \sum\limits_{j\in \mathcal{L} \backslash \{i\}}\theta_{j,m,(k,t)}  P^{\rm V}_{j,m,(k,t)} G_{j,i,m,(k,t)}.\IEEEnonumber
    \end{align}
    
    We discretize the transmit power $P^{\rm V}_{i,m,(k,t)}$ to four levels for ease of learning and practical circuit restriction \cite{liang2019spectrum,Xiang2021MARLDecentralizedSA}, i.e., $P^{\rm V}_{i,m,(k,t)}\in\mathcal{A}_P=\{23, 10, 5, -100\}$ $\rm dBm$. Note that $-100$ dBm can be considered as zero transmit power. \par
    
    \subsubsection{Instantaneous Data Rate}
    
    The instantaneous data rates $r_{m,(k,t)}$ and $r_{i,(k,t)}$ of V2I link $m$ and V2V link $i$ at communication interval $(k,t)$ are respectively derived as
    \begin{equation}
    \label{rateV2I}
    r_{m,(k,t)} = W\log_2(1+\gamma_{m,(k,t)}),
    \end{equation}
    and
    \begin{align}\label{rateV2V}
    r_{i,(k,t)} =\sum_{m=0}^{M-1} {\theta_{i,m,(k,t)}} W\log_2(1+\gamma_{i,m,(k,t)}),
    \end{align}
    \noindent where $W$ is the bandwidth of a subchannel.\par Let $r^{\rm CAM}_{i,(k,t)}$ denote the transmission rate of V2V link $i$ in terms of CAM at communication interval $(k,t)$, which is given by
    \begin{equation}
    \label{rateV2VCAM}
    r^{\rm CAM}_{i,(k,t)}=\frac{r_{i,(k,t)}}{N_c},
    \end{equation}
    \noindent where $N_c$ is the constant CAM size. \par
    
    \subsubsection{Queuing Dynamic}
    Each V2V transmitter $i\in\mathcal{L}$ has a buffer to store its CAM. Let $q^{\rm CAM}_{i,(k,t)}$ denote the queue length of V2V transmitter $i$, in number of CAMs, at communication interval $(k,t)$. At the beginning of the first communication interval $(k, 0)$ of each control interval $k\in\mathcal{K}$, V2V transmitter $i$ samples the driving status and generates a new CAM, which replaces any CAMs left undelivered from the previous control interval. Therefore, the queue length at the start of each control interval is set to $q^{\rm CAM}_{i,(k,0)}=1$ \cite{lei2023multitimescale}. \par At each communication interval $(k, t), t\in\mathcal{T}$, within control interval $k$, the queue length decreases according to the number of CAMs transmitted, with the constraint that it cannot become negative: $q^{\rm CAM}_{i,(k,t+1)}=\max \left[0,q^{\rm CAM}_{i,(k,t)}- r^{\rm CAM}_{i,(k,t)}\Delta t\right]$. \par
    
    To capture the discontinuity at control interval boundaries, we denote by $q^{\rm CAM}_{i,(k,T)}$ the queue length at the end of the last communication interval $(k,T-1)$ of control interval $k$. Note that $q^{\rm CAM}_{i,(k,T)}\neq q^{\rm CAM}_{i,(k+1,0)}$ because the queue is reset at the start of each new control interval.\par

       The queue process thus evolves as 
\begin{align}
\label{queue2}
\setlength{\arraycolsep}{1.6pt}
& q^{\rm CAM}_{i,(k,t)}
= \IEEEnonumber \\ 
&\left\{
\begin{array}{ll}
1, & \mathrm{if}\quad t=0, \\[1mm]
\max\left[0,q^{\rm CAM}_{i,(k,t-1)}
-r^{\rm CAM}_{i,(k,t-1)}\Delta t\right],
& \mathrm{if}\quad t=1,\ldots,T.
\end{array}
\right.
\end{align}

 	\section{C-V2X Multi-agent Environment}
 	
 	In this section, we formulate the C-V2X RRA problem
as a hierarchy of multi-agent interference games.
Each successive game incorporates additional
environmental features that introduce new MARL
challenges and may interact with those already present. \par
  
  %By comparing the performance of various MARL algorithms under these games, we aim to reveal insights into which algorithms work better and why. \par 
 	
 %	When formulating the games, only large-scale fading is taken into account when calculating the channel gains. We will demonstrate that including the small-scale fading effects does not alter the nature of these games, although the additional stochasticity poses more challenges for the convergence of MARL algorithms. \par   
 	
 	\subsection{Normal-Form Interference Game (NFIG)}
  \subsubsection{Environment model}
 	Without loss of generality, we consider a snapshot of the channel gains at the communication interval $(k,t)$ is captured. The objective is to derive the optimal policy that prescribes the subchannel allocation $\{\theta_{i,m,(k,t)}\}_{m\in\mathcal{M}}$ and transmit power $\{P^{\rm V}_{i,m,(k,t)}\}_{m\in\mathcal{M}}$ for each V2V link $i$, so that the sum data rate of all the V2V and V2I links is maximized for this specific communication interval. This simple problem can be formulated as a normal-form game consisting of

\begin{itemize}
    \item Finite set of V2V agents $\mathcal{L}$;
    \item For each agent $i\in\mathcal{L}$:
    \begin{itemize}
        \item action $a_{i}=\{\theta_{i,m},P^{\rm V}_{i,m}\}_{m\in\mathcal{M}}$;
        \item common reward 
        \begin{align}
        \label{NFIG_reward}
        R = \lambda_1 \sum_{m\in\mathcal{M}} r_{m,(k,t)} + \lambda_2 \sum_{i\in\mathcal{L}} r_{i,(k,t)},
        \end{align}
        where $\lambda_1$ and $\lambda_2$ are weighting coefficients reflecting the relative importance of V2I and V2V objectives, respectively. 
    \end{itemize}
\end{itemize}
    
    Since the normal-form game defines a single interaction between the agents and environment, i.e., the time horizon is $1$, the communication interval index $(k,t)$ is omitted in the above notations. Each agent $i$ samples an action $a_{i}$ with probability $\pi_{i}(a_{i})$ given by its policy. The actions of all the agents form a joint action $a=\{a_{i}\}_{i\in\mathcal{L}}$. Finally, each agent receives the common reward based on the reward function and the joint action. \par
 	
%      Fig. \ref{matrix game} illustrates a simple example where two V2V links share a single subchannel that is not occupied by any V2I link. For notational simplicity, we omit the subchannel index. Each agent only has two possible actions in the action space: 
%  \begin{itemize}
%  	\item action $\rm{T}$: transmit at maximum power, i.e., $a_{i}=\{\theta_{i}=1,P^{\rm V}_{i}=23 \rm{dBm}\}$;
%  	\item action $\rm{NT}$: do not transmit, i.e., $a_{i}=\{\theta_{i}=0,P^{\rm V}_{i}=-100\rm{dBm}\}$.
%  \end{itemize}
%  
%  \begin{figure}
%  	\centering
%  	\includegraphics[width=0.5\textwidth]{game.pdf}
%  	\caption{A simple normal-form interference game for V2V communications }
%  	\label{matrix game}
%  \end{figure} 	

 \subsubsection{MARL challenges}
The main challenges faced by MARL algorithms in NFIG
are coordination difficulty and non-stationarity.
Coordination difficulty arises from the interdependence
of agents' actions and rewards, particularly when agents
lack knowledge of other agents' actions or policies.
During exploration, an agent may receive a low reward
due to poor decisions made by other agents, even when
its own action is part of an optimal joint action.
When miscoordination penalties are high, agents may
favor ``safe'' actions that provide acceptable rewards
regardless of the actions of others. This can lead to
two well-known coordination pathologies
\cite{JMLR:v17:15-417}:

\begin{itemize}
    \item \emph{Relative overgeneralization}:
    A suboptimal Nash equilibrium may be preferred
    over an optimal Nash equilibrium because the
    actions associated with the former yield higher
    expected rewards when averaged over other
    agents' exploratory actions.

    \item \emph{Miscoordination}:
    When multiple optimal Nash equilibria exist,
    agents may select actions associated with
    different equilibria, resulting in a suboptimal
    joint action.
\end{itemize}

Another fundamental MARL challenge is
\emph{non-stationarity}, arising from the evolving
policies of other agents during simultaneous learning.
These policy changes alter the effective environment
perceived by each agent, causing action values to
change over time and making stable learning and
convergence more difficult.
   
    \subsection{Stochastic Interference Game (SIG)}
    \subsubsection{Environment model}
    Consider the more practical case when the instantaneous data rate of the V2V links and V2I links vary over time due to the changing distances between the transmitters and receivers and the fast fading effects. The optimization objective is to maximize the V2I throughput as well as the probability of fully delivering the newly generated CAM within each control interval. This problem can be formulated as a stochastic game consisting of 
      \begin{itemize}
     	\item State $s_{(k,t)}=(G_{(k,t)},q_{(k,t)}^{\mathrm{CAM}},t)$, where $G_{(k,t)}=\{\{G_{i,m,(k,t)}\}_{i\in\mathcal{L}},\{G_{j,i,m,(k,t)}\}_{i,j\in\mathcal{L}},\{G_{B,i,m,(k,t)}\}_{i\in\mathcal{L}},$ $\{G_{i,B,m,(k,t)}\}_{i\in\mathcal{L}},G_{m,(k,t)}\}_{m\in\mathcal{M}}$, and $q_{(k,t)}^{\mathrm{CAM}}=\{q_{i,(k,t)}^{\mathrm{CAM}}\}_{i\in\mathcal{L}}$. The state is thus composed of the channel gains of all the V2V links, V2I links, and interference links on all the subchannels, as well as the queue length of all the V2V links. 
        % \item Common reward    	

        %     \begin{align}
        %     \label{GR}
        %     R_{(k,t)}=&\lambda_1 \sum_{m\in \mathcal{M}} r_{m,(k,t)}+ \IEEEnonumber \\
        %     &\lambda_2 \sum_{i\in \mathcal{L}}r_{i,(k,t)}|_{q^{\rm CAM}_{i,(k,t)}>0}+ \sum_{i\in \mathcal{L}}Z|_{q^{\rm CAM}_{i,(k,t)}=0},
        %     \end{align}

        \item Common reward
        \begin{equation}
        \label{GR}
        \begin{split}
        R_{(k,t)} = &\lambda_1 \sum_{m\in \mathcal{M}} r_{m,(k,t)} \\
                    &+ \lambda_2 \sum_{i\in \mathcal{L}} r_{i,(k,t)}|_{q^{\mathrm{CAM}}_{i,(k,t)}>0}
                     + \sum_{i\in \mathcal{L}} Z|_{q^{\mathrm{CAM}}_{i,(k,t)}=0},
        \end{split}
        \end{equation}
            
            \noindent where $Z$ is a tuned hyper-parameter encouraging early completion of transmission for CAM whose value is greater than the largest $r_{i,(k,t)}$ ever obtained \cite{liang2019spectrum}. Note that $Z$ is granted in each communication interval for which the queue is empty, rather than only once upon completion of CAM transmission.
            
      %We set $G=10W$ in the experiments.  
    	\item State transition probability 
    	\begin{align}
    	&\mathrm{Pr}(s_{(k,t+1)}| s_{(k,t)},a_{(k,t)})=\IEEEnonumber \\
     &\mathrm{Pr}(G_{(k,t+1)}| G_{(k,t)}) 
        \mathrm{Pr}(q^{\rm CAM}_{(k,t+1)}|q^{\rm CAM}_{(k,t)},G_{(k,t)},a_{(k,t)}),
    	\end{align}
    	\noindent where $\mathrm{Pr}(q^{\rm CAM}_{(k,t+1)}|q^{\rm CAM}_{(k,t)},G_{(k,t)},a_{(k,t)})$ can be derived from \eqref{queue2}. 
    \end{itemize}
%Note that the state transition probability depends on the action $a_{(k,t)}$, which have long-term consequences within the horizon of a single control interval. 

Note that, at the beginning of each control interval $k$, the queue length $q^{\rm CAM}_{(k,0)}$ is reset to $1$, and the actions taken during the past control intervals $k'<k$ do not affect the queue length at control interval $k$. Therefore, the time horizon of SIG is $T$ communication intervals in a single control interval.\par

\subsubsection{MARL challenges}
SIG retains the coordination difficulty and
non-stationarity present in NFIG while introducing
additional challenges associated with its
environmental features.

First, multiple time steps introduce the challenge
of sequential decision making, as resource allocation
decisions affect subsequent queue states and future
rewards. The consequences of suboptimal joint actions
may therefore propagate across multiple time steps,
further complicating coordination and policy learning.

Second, fast fading introduces environmental
stochasticity through time-varying channel conditions
within an episode. Agents must learn effective
resource allocation policies under stochastic
channel variations. Although fast fading does not
necessarily introduce non-stationarity, it may
increase learning difficulty and interact with
existing MARL challenges.

Third, increasing the number of agents enlarges
the joint action space and may intensify coordination
difficulty and non-stationarity.

Finally, multiple vehicular topologies introduce
diverse path-loss and interference conditions across
episodes. Learning a common policy across these
topologies introduces the challenge of robustness
over the training topologies and generalization
to unseen topologies \cite{pmlr-v119-cobbe20a}.

    \subsection{Partially Observable Stochastic Interference Game (POSIG)}
 \subsubsection{Environment model}
    In SIG, it is assumed that each agent $i$ can observe the global state $s_{(k,t)}$ in each communication interval $(k,t)$. This assumption is normally not realistic in practical scenarios. Instead, each agent $i$ can only receive local observations consisting of partial information about the state $s_{(k,t)}$. This scenario can be formulated as a POSIG, which is defined by the same elements of SIG. Additionally, the observation for each agent $i\in\mathcal{L}$ is defined as \cite{ye2019deep}:
    \begin{align}
    s_{i,(k,t)}=\{ G_{i,(k,t)},I_{i,(k,t-1)},q^{\rm CAM}_{i,(k,t)},t\},
    \end{align}
   \noindent where $G_{i,(k,t)}=\{G_{i,m,(k,t)},G_{i,B,m,(k,t)}\}_{m\in\mathcal{M}}$ is the local observation of channel states at V2V transmitter $i$ in communication interval $(k,t)$. Moreover, $I_{i,(k,t-1)}=\{I_{i,m,(k,t-1)}\}_{m\in\mathcal{M}}$ is the received interference power at V2V receiver $i$ over all subchannels at the previous communication interval $(k,t-1)$.

 \subsubsection{MARL challenges}
POSIG restricts agents to local observations,
introducing partial observability in addition to
the challenges present in SIG. Different global
states may produce the same local observation,
preventing an agent from fully distinguishing
the underlying interference conditions when
selecting its action. This information limitation
may further complicate coordination and policy
learning.

\begin{comment}
{\color{blue} 
\begin{table*}[h]
\centering
\caption{Overview of Learning Tasks and MARL Challenges}
\label{tab:tasks_challenges}
\begin{tabular}{@{}lccccccc@{}}
\toprule
\textbf{Task} & \textbf{Horizon} & \textbf{Coord.} & \textbf{Non-stat.} & \textbf{Large Action Space} & \textbf{Adapt./General.} & \textbf{Partial Obs.} \\
\midrule
\emph{NFIG} & 1 & \checkmark & \checkmark & - & - & - \\
\emph{SIG SL\_NFF} & 50 & \checkmark & \checkmark & - & - & - \\
\emph{SIG SL\_FF} & 50 & \checkmark & \checkmark & \checkmark\,($L{>}4$) & - & - \\
\emph{SIG ML} & 50 & \checkmark & \checkmark & \checkmark\,($L{>}4$) & \checkmark & - \\
\emph{POSIG} & 50 & \checkmark & \checkmark & \checkmark\,($L{>}4$) & \checkmark & \checkmark \\
\bottomrule
\end{tabular}
\end{table*}
}
\end{comment}

\section{Algorithms}
We evaluate eight representative MARL algorithms spanning value-based and actor-critic methods under IL and CTDE. Their main characteristics are summarized in Table~\ref{tab:algorithm_summary}.
 \subsection{Independent Learning (IL)}
In IL, each agent treats the other agents as part of the environment and uses single-agent DRL algorithms to learn a policy. 

\subsubsection{Value-based}
\begin{description}
  	\item[IDQN] In Independent DQN (IDQN), each agent has a pair of Q-network and target network and uses the off-policy DQN \cite{mnih2015human} for decentralized training in a multi-agent environment. 
    
    \item[Hys-IDQN] Hysteretic IDQN (Hys-IDQN) is a variant of IDQN aimed at enhancing coordination in IL. Unlike standard IDQN, it employs two distinct learning rates based on the sign of the Temporal-Difference (TD) error \cite{matignon2007hysteretic}.
\end{description}

\subsubsection{Actor-critic}
\begin{description}
  	\item[IA2C] Advantage Actor-Critic (A2C)~\cite{mnih2016asynchronous} is an on-policy actor-critic algorithm. Its independent, per-agent variant, IA2C, assigns each agent a pair of actor and critic networks for decentralized training in a multi-agent environment.
    
    \item[IPPO] In IPPO, each agent uses the on-policy PPO \cite{schulman2017proximal} for decentralized training in a multi-agent environment. Compared to IA2C, IPPO optimizes a surrogate objective by performing multiple update epochs within each training iteration. 
\end{description}

\subsection{Centralized Training with Decentralized Execution (CTDE)}
Compared with IL, CTDE uses centralized information and joint learning structures during training to facilitate coordination or improve value estimation.

\subsubsection{Value Decomposition}
In value-based CTDE, the joint Q-function is represented in terms of individual utility functions learned during centralized training. Ideally, the Individual-Global-Max (IGM) condition should be satisfied, such that the joint action obtained by individually maximizing the agents' utility functions also maximizes the joint Q-function. During decentralized execution, each agent independently selects its action according to its individual utility function.

  \begin{description}

  	\item[VDN] In Value Decomposition Networks (VDN) \cite{sunehag2017value}, the joint Q-function is the sum of individual utility functions.  
    \item[QMIX] In QMIX \cite{rashid2018qmix}, the joint Q-function is a monotonic function of the individual utility functions. The monotonicity constraint is more general than the additive factorization of VDN while preserving the IGM property.     

   \end{description}
 
 \subsubsection{Centralized Critic Decentralized Actors}
In actor-critic CTDE, a centralized critic learns a value function using global information during training, while decentralized actors learn individual policies for execution. In partially observable environments, the actors are conditioned on locally available observations or observation-action histories.
 
\begin{description}
  	\item[MAA2C] Multi-agent A2C (MAA2C) \cite{papoudakis2021benchmarking} differs from IA2C by incorporating the global state into the critic, alongside local observations and previous actions, in environments with partial observability. 
    \item[MAPPO] The relationship between Multi-agent PPO (MAPPO) \cite{yu2022surprising} and IPPO mirrors that between MAA2C and IA2C.
\end{description}

The MADDPG algorithm \cite{lowe2020multi}, another well-known actor-critic CTDE method, is excluded because it is primarily designed for continuous action spaces, whereas our C-V2X tasks involve discrete actions. Moreover, \cite{papoudakis2021benchmarking} shows that MADDPG underperforms relative to other algorithms across most evaluated games. \par

\subsection{Comparison of MARL Paradigms}

The above algorithms differ in their learning and coordination mechanisms, which may affect their ability to address the MARL challenges considered in this work.

\subsubsection{Value-Based vs. Actor-Critic Learning}
The value-based methods learn action values through off-policy TD learning with experience replay and bootstrapped targets. In multi-agent learning, the effective environment changes as the policies of other agents evolve, so replayed transitions generated under earlier joint policies can become increasingly mismatched with the current learning environment. Moreover, value estimation errors can propagate through successive bootstrapped TD targets, potentially making learning more difficult as the scale and complexity of the problem increase.\par

The evaluated actor-critic methods instead directly optimize parameterized policies using on-policy trajectories. Their value estimates can exploit multi-step returns, reducing their reliance on short-horizon bootstrapping, while their on-policy updates avoid the additional mismatch introduced by replaying experience from earlier joint policies. PPO further constrains policy changes through its clipped surrogate objective, which can improve training stability.

\subsubsection{IL vs. CTDE}
Compared with value-based IL, value-based CTDE uses value decomposition to facilitate coordinated learning. However, the IGM requirement constrains the joint action dependencies that can be represented. In particular, the additive factorization of VDN and monotonic factorization of QMIX prevent an agent's action preference from being reversed by the actions of other agents. Such preference reversals can arise naturally in interference-coupled RRA, where the value of selecting a resource depends strongly on the resources selected by other agents. Value-based CTDE may therefore improve coordination over value-based IL, but remains limited when coordination requires such complex joint action dependencies.

Compared with actor-critic IL, actor-critic CTDE uses a centralized critic that, under partial observability, can exploit global information unavailable to independent critics, potentially improving value estimation and policy learning. However, this advantage is limited by the centralized--decentralized mismatch, where the centralized critic uses global information to guide policy learning, while the decentralized actors must realize the resulting coordination using only local information during execution. Thus, improved centralized value estimation does not necessarily translate into effective decentralized coordination. In fully observable tasks, this information advantage largely disappears because independent critics also have access to the full global-state information. 

\begin{table}[h]
\centering
\caption{Summary of Evaluated MARL Algorithms}
\label{tab:algorithm_summary}
\small
\begin{tabular}{@{}llcc@{}}
\toprule
\textbf{Algorithm} & \textbf{Type} & \textbf{Framework} & \textbf{On/Off-Policy} \\
\midrule
IDQN & Value-based & IL & Off-policy \\
Hys-IDQN & Value-based & IL & Off-policy \\
VDN & Value-based & CTDE & Off-policy \\
QMIX & Value-based & CTDE & Off-policy \\
\midrule
IA2C & Actor-critic & IL & On-policy \\
IPPO & Actor-critic & IL & On-policy \\
MAA2C & Actor-critic & CTDE & On-policy \\
MAPPO & Actor-critic & CTDE & On-policy \\
\bottomrule
\end{tabular}
\end{table}

\section{Experiment Methodology}  	
We set up the simulated environment based on the evaluation methodology for the freeway case described in Annex A of 3GPP TR~36.885~\cite{3gpp_release14}, including the road configuration, V2V and V2I channel models, and the V2V traffic model. For vehicle density and speed, we adopt three representative scenarios from ETSI TR 103 766 \cite{ETSITR103766}: Scenario \#1 (35 veh/km, 250 km/h), \#3 (123 veh/km, 70 km/h), and \#6 (500 veh/km, 50 km/h). The key parameters of the C-V2X environment are summarized in Table~\ref{tab:technical_params}, following existing works \cite{3gpp_release14,liang2019spectrum}. \par 

For the reward functions, we set the weights with $\lambda_1{:}\lambda_2 = 1{:}9$ to prioritize CAM delivery over V2I throughput, reflecting the heterogeneous quality-of-service (QoS) requirements of C-V2X systems and following the V2V-favoring weighting commonly adopted in C-V2X resource allocation studies~\cite{ye2019deep,10118603,Ding2022AMARL,xu2024federated}. Following the reward-shaping principle in~\cite{liang2019spectrum}, we choose $Z$ to be larger than the maximum achievable single-step V2V transmission rate, so that completing a V2V transmission is always preferred to any single-step throughput gain, and empirically tune it within the range suggested in~\cite{liang2019spectrum}. This yields $Z = 0.5$ for \emph{SIG}/\emph{POSIG}.
% TABLE II (Vehicle Densities and Speeds) - DELETED, content inlined above
\begin{table}[h]
\centering
\caption{Parameters of C-V2X Environment}
\label{tab:technical_params}
\begin{tabular}{ll} 
\toprule
\textbf{Description}           & \textbf{Value}            \\ 
\midrule
Communication interval         & $1$ ms                      \\ 
Control interval & $100$ ms             \\ 
Carrier frequency & $2$ GHz              \\ 
Total Bandwidth       & $4$ MHz                   \\ 
V2I link/subchannel number &  $4$  \\
V2V link/agent number & $\{4, 8, 16\}$   \\
Noise power                    & $-114$ dBm                  \\ 
V2I transmit power             & $23$ dBm   \\ 
V2V transmit power             & $\{23, 10, 5, -100\}$ dBm  \\
BS antenna height              & $25$ m \\
BS antenna gain             & $8$ dBi \\
BS noise figure             & $5$ dB \\
Vehicle antenna gain        & $3$ dBi \\
Vehicle noise figure        & $9$ dB \\
CAM size                             & $6\times1{,}060$ bytes \\
\bottomrule
\end{tabular}
\end{table}
\subsection{Vehicle Positional Data Generation}
We simulate a 2 km highway with three lanes in each direction (six lanes in total). Vehicle positional data is obtained from SUMO, an open-source traffic simulation platform that provides realistic vehicle movement data. We create datasets for all tasks with the goal of representing diverse interference conditions. \par

To capture realistic interference conditions while keeping computational complexity manageable, we adopt the following procedure. Initial topologies are generated by randomly placing the vehicles according to a spatial Poisson process at one end of the 2 km highway. All vehicles are confined within a distance of $D$ m from the end of the highway, where $D$ is given by the ratio of the total number of vehicles to the vehicle density. Vehicles then travel at time-varying speeds with random lane changes, and positional data is sampled every 100 ms. This procedure is repeated under the three vehicle-density levels. \par

The generated topologies are partitioned into training and held-out sets, so that both sets follow the same data-generation process. The training dataset contains samples evenly distributed across the three density levels. We use 15,000 training topologies for the 4-agent setting and 60,000 for the 8- and 16-agent settings. \par

From the held-out data, we select nine evaluation topologies covering the combinations of three vehicle-density levels (35, 123, and 500~veh/km) and three vehicle-to-BS distance levels (\texttt{close}, \texttt{mid}, and \texttt{far}), forming a $3\times3$ evaluation matrix. The same nine evaluation topologies are used across the hierarchy of learning tasks, enabling controlled comparisons under a common set of interference conditions and, for the topology-specific \emph{NFIG} and \emph{SIG SL} tasks, comparison across different representative topologies. The number of subchannels is fixed at $M=4$, while the number of agents varies across $L\in\{4,8,16\}$.\par

\subsection{Learning Tasks}
Table~\ref{tab:tasks} summarizes the benchmark learning
tasks, the additional features introduced at each stage,
and their corresponding MARL challenges. 

\subsubsection{NFIG}
\emph{NFIG} is the simplest task, where agents learn
resource allocation for a fixed vehicle topology in a
single time step. Simultaneous multi-agent learning
introduces coordination difficulty and non-stationarity,
which may cause learning to converge to a suboptimal
Nash equilibrium, as discussed in Section~IV.A.2).
We consider nine evaluation topologies, each trained
and evaluated independently with $L=4$ agents.
This small scale enables exhaustive search for the
globally optimal joint action, allowing the learned
performance to be directly compared with the optimum.

\subsubsection{SIG}
Compared with \emph{NFIG}, \emph{SIG} incorporates
additional environmental features, including multiple
time steps, fast fading, and varying vehicular topologies.
We construct a sequence of \emph{SIG}-based learning
tasks that progressively introduces these features,
together with increasing network scale, to examine
their associated MARL challenges and effects on
learning performance.

\paragraph{SIG Single Location with No Fast Fading (SL\_NFF)}
\emph{SIG SL\_NFF} uses a single fixed topology without
fast fading. Compared with \emph{NFIG}, it introduces
multiple time steps, giving rise to the challenge of
sequential decision making, while retaining the same
topology and deterministic channel evolution.
We set the CAM size to $N_c=6\times1{,}060$ bytes
and use a 50-ms training horizon because the queue
is typically depleted within the first 20 time steps
under a reasonable policy; the learned policies remain
applicable to the full 100-ms horizon.

\paragraph{SIG Single Location with Fast Fading (SL\_FF)}
\emph{SIG SL\_FF} further introduces fast fading,
giving rise to environmental stochasticity through
time-varying channel conditions within each episode.
We additionally vary the number of simultaneously
interacting agents as $L\in\{4,8,16\}$, enlarging
the joint action space and potentially intensifying
coordination difficulty and non-stationarity.

\paragraph{SIG Multiple Locations (ML)}
\emph{SIG ML} trains a common policy across diverse
vehicular topologies in the training dataset and
evaluates it on the nine evaluation topologies
described in Section~VI.A. Compared with
\emph{SIG SL\_FF}, it introduces multiple vehicular
topologies, giving rise to the challenges of policy
robustness across varying interference conditions
and generalization to unseen topologies.

For \emph{SIG ML}, we found that global-state ordering
can substantially affect learning performance despite
unchanged information content. We therefore use an
agent-specific ordering for IL and decentralized actors,
placing each agent's own link information first,
followed by that of the other agents. This provides
a consistent agent-centered representation across
diverse topologies. The centralized critic instead
uses the original unordered global state, which is
shared across all agents.

\subsubsection{POSIG}
\emph{POSIG} uses the same topology datasets and
environmental configurations as \emph{SIG ML}, but
restricts each agent to locally available channel
and queue information. This additional information
constraint introduces partial observability.
Together with the controlled state-representation
comparisons described in Section~VII, this setting
is used to examine its effect on learning performance.

\begin{table}[t]

\centering
\caption{Additional task features and their corresponding
MARL challenges across the benchmark hierarchy.}
\label{tab:tasks}
\footnotesize
\setlength{\tabcolsep}{3pt}
\renewcommand{\arraystretch}{1.2}
\begin{tabular}{@{}lll@{}}
\toprule
\textbf{Task}
& \makecell[l]{\textbf{Additional}\\\textbf{task feature}}
& \makecell[l]{\textbf{Principal additional}\\\textbf{MARL challenge}} \\
\midrule
\emph{NFIG}
& Simultaneous learning
& Coordination, non-stationarity \\
\emph{SIG~SL\_NFF}
& Multiple time steps
& Sequential decision making \\
\emph{SIG~SL\_FF}
& Fast fading
& Environmental stochasticity \\
\emph{SIG~SL\_FF}
& More agents
& Larger joint action space \\
\emph{SIG~ML}
& Multiple topologies
& Robustness, generalization \\
\emph{POSIG}
& Local observations
& Partial observability \\
\bottomrule
\end{tabular}
\end{table}

\begin{comment}
\begin{table*}[t]
\centering
\small
\caption{Overview of Properties and MARL Challenges of Learning Tasks. \emph{FF} denotes Fast Fading. \emph{Non-Stat.} denotes Non-Stationarity. \emph{Robust./Gen.} denotes Robustness/Generalization.}
\label{tab:marl_tasks}
\setlength{\tabcolsep}{4pt}
\begin{tabular}{lccc|ccccc}
\hline
\multirow{2}{*}{Task}
& \multicolumn{3}{c|}{Properties}
& \multicolumn{5}{c}{MARL Challenges} \\
\cline{2-9}
& Steps & Agents & FF
& Non-Stat.
& Coordination
& Large Action Space
& Robust./Gen.
& Partial Observability \\
\hline
\emph{NFIG}
& 1 & 4 & 
& \checkmark & \checkmark & & & \\
\emph{SIG SL\_NFF}
& 50 & 4 &
& \checkmark & \checkmark & & & \\
\emph{SIG SL\_FF}
& 50 & 4,8,16
& \checkmark & \checkmark & \checkmark & \checkmark & & \\
\emph{SIG ML}
& 50 & 4,8,16
& \checkmark & \checkmark & \checkmark & \checkmark & \checkmark & \\
\emph{POSIG}
& 50 & 4,8,16
& \checkmark & \checkmark & \checkmark & \checkmark & \checkmark & \checkmark \\
\hline
\end{tabular}
\end{table*}
\end{comment}

\subsection{Evaluation Protocol}

Each algorithm was run on all five learning tasks (Table~\ref{tab:tasks}), using five random seeds for each task. We train all algorithms for a number of environment time steps, evaluating performance at regular intervals throughout training, resulting in a total of 100 evaluations per run. At each evaluation point, nine evaluation episodes were executed without exploration noise, and the average return across these episodes was calculated. The nine evaluation episodes correspond to the nine evaluation topologies for \emph{SIG ML} and \emph{POSIG} tasks, while they are conducted on the single training topology for other tasks. For each algorithm, we identify the evaluation point at which the average return across the five independent runs is maximized. The corresponding policy from each run is retained for subsequent evaluation. \par

To ensure a fair comparison, we account for the inherent differences in sample efficiency between off-policy value-based and on-policy actor-critic algorithms when setting the training horizon ~\cite{papoudakis2021benchmarking}. For each task and algorithm category, we empirically select the number of environment time steps based on the observed learning curves, aiming to ensure that the algorithms approach stable performance within their allocated training horizons. Accordingly, all algorithms interact with the environment for $0.05$ million time steps in \emph{NFIG}. In \emph{SIG SL}, the value-based and actor-critic methods interact with the environment for $0.75$ and $5$ million time steps, respectively, while in \emph{SIG ML} and \emph{POSIG}, they interact for $1.5$ and $10$ million time steps, respectively. These time-step budgets specify the amount of experience collected rather than the number of parameter updates, which differs across algorithms according to their respective update schemes. Although the actor-critic methods use approximately $6.7\times$ more environment time steps, their computational costs are not systematically higher. For example, in the 16-agent \emph{SIG ML} task, the value-based and actor-critic methods require 17.5--23.9 and 13.1--26.4 CPU core-hours per run, respectively, under the same hardware configuration. Thus, the different training horizons account for differences in sample efficiency and convergence without imposing systematically higher computational costs on the actor-critic methods.

Hyperparameters were tuned for each task category by evaluating each configuration over five seeds and selecting the best-performing one. The final values used for training the different algorithms are summarized in the Appendix.

\subsection{Performance Metrics}
To facilitate comparison across algorithms and tasks with different reward scales, we normalize the return as
\[
\mathrm{norm}\_G_t^a= \frac{G_t^a-G_t^{\min}} {G_t^{\max}-G_t^{\min}},
\]
where $G_t^{\min}$ and $G_t^{\max}$ are algorithm-independent low- and high-performance references, respectively. For all tasks, $G_t^{\min}$ is obtained using a random policy. The high-performance reference $G_t^{\max}$ is determined as follows.

\textbf{\emph{NFIG}:} Since each episode contains a single time step with four agents, exhaustive search over the joint action space is feasible, and $G_t^{\max}$ is the globally optimal return.

\textbf{\emph{SIG SL}:} For four agents, $G_t^{\max}$ is obtained by per-step exhaustive search that maximizes the instantaneous common reward in \eqref{GR}. This provides a high-performance reference that tends to empty the V2V queues as quickly as possible, thereby increasing the likelihood of obtaining the early-completion bonus \mbox{$Z$}. For 8 and 16 agents, where exhaustive search is computationally prohibitive, we use a greedy assignment that, at each step, allocates distinct subchannels to the agents in order of priority and then lets each agent in turn switch to the subchannel--power pair that most increases the instantaneous common reward. For four agents, this approximation achieves an average return only $1.9\%$ below per-step exhaustive search.

\textbf{\emph{SIG ML} and \emph{POSIG}:} For each of the nine evaluation topologies, we reuse the corresponding \(G_t^{\max}\) obtained for \emph{SIG SL}.

Except in \emph{NFIG}, $G_t^{\max}$ is a reference baseline rather than the optimal performance, so normalized returns may occasionally exceed one. Because the same algorithm-independent references are applied to all algorithms within each task, they do not affect the relative algorithm comparisons.

For each algorithm, we report the maximum average evaluation return achieved during training, where the average is computed over five independent runs with different random seeds. Table~\ref{tab:benchmarking} presents these maximum returns along with their 95\% confidence intervals. The top-performing algorithm in each task is indicated in bold.  \par

Since repeatedly evaluating on the same nine topologies when determining the maximum return in \emph{SIG ML} and \emph{POSIG} may introduce optimistic bias, we additionally evaluate the corresponding policies from the five runs on 100 held-out topologies sampled from the same data-generation process across the three vehicle-density levels. The normalized returns differ from those on the nine evaluation topologies by at most $0.06$ and by $0.02$ on average. The held-out returns are generally slightly lower, indicating a small optimistic bias, while the main comparative conclusions remain unchanged. \par

\newcommand{\rot}[1]{\rotatebox[origin=c]{90}{#1}}
\begin{table*}[ht]
  \centering
  \caption{Maximum Normalized Returns and 95\% Confidence Intervals over Five Seeds Across All Tasks}
  \label{tab:tasks_summary}
  \footnotesize
  \begin{tabular}{@{}llcccccccc@{}}
    \toprule
    \textbf{} & \textbf{Tasks/Algos.} & \textbf{IDQN} & \textbf{Hys-IDQN} & \textbf{VDN} & \textbf{QMIX} & \textbf{IA2C} & \textbf{MAA2C} & \textbf{IPPO} & \textbf{MAPPO} \\
    \midrule
    
\multirow{11}{*}{\rotatebox{90}{\emph{NFIG}}} 
    & $35\_\text{far}$    & $\bm{1.00 \pm 0.01}$ & $0.97 \pm 0.07$ & $\bm{1.00 \pm 0.01}$ & $\bm{1.00 \pm 0.01}$ & $0.91 \pm 0.09$ & $0.93 \pm 0.09$ & $0.87 \pm 0.08$ & $0.87 \pm 0.09$ \\
    & $35\_\text{mid}$    & $\bm{1.00 \pm 0.00}$ & $\bm{1.00 \pm 0.00}$ & $\bm{1.00 \pm 0.00}$ & $\bm{1.00 \pm 0.00}$ & $\bm{1.00 \pm 0.00}$ & $\bm{1.00 \pm 0.00}$ & $\bm{1.00 \pm 0.00}$ & $\bm{1.00 \pm 0.00}$ \\
    & $35\_\text{close}$  & $\bm{1.00 \pm 0.01}$ & $\bm{1.00 \pm 0.01}$ & $\bm{1.00 \pm 0.00}$ & $\bm{1.00 \pm 0.00}$ & $\bm{1.00 \pm 0.00}$ & $\bm{1.00 \pm 0.00}$ & $0.99 \pm 0.01$ & $0.99 \pm 0.01$ \\
    & $123\_\text{far}$   & $\bm{1.00 \pm 0.00}$ & $0.99 \pm 0.01$ & $\bm{1.00 \pm 0.00}$ & $\bm{1.00 \pm 0.00}$ & $\bm{1.00 \pm 0.00}$ & $\bm{1.00 \pm 0.01}$ & $\bm{1.00 \pm 0.00}$ & $\bm{1.00 \pm 0.00}$ \\
    & $123\_\text{mid}$   & $\bm{0.99 \pm 0.02}$ & $0.98 \pm 0.02$ & $\bm{0.99 \pm 0.02}$ & $\bm{0.99 \pm 0.01}$ & $\bm{0.99 \pm 0.04}$ & $0.97 \pm 0.03$ & $0.97 \pm 0.01$ & $0.97 \pm 0.00$ \\
    & $123\_\text{close}$ & $0.99 \pm 0.01$ & $\bm{1.00 \pm 0.00}$ & $\bm{1.00 \pm 0.00}$ & $\bm{1.00 \pm 0.00}$ & $0.99 \pm 0.00$ & $0.99 \pm 0.00$ & $0.99 \pm 0.00$ & $0.99 \pm 0.02$ \\
    & $500\_\text{far}$   & $0.99 \pm 0.01$ & $\bm{1.00 \pm 0.01}$ & $\bm{1.00 \pm 0.00}$ & $\bm{1.00 \pm 0.00}$ & $0.93 \pm 0.05$ & $0.92 \pm 0.01$ & $0.99 \pm 0.01$ & $0.99 \pm 0.01$ \\
    & $500\_\text{mid}$   & $\bm{1.00 \pm 0.00}$ & $0.98 \pm 0.05$ & $\bm{1.00 \pm 0.00}$ & $\bm{1.00 \pm 0.00}$ & $0.98 \pm 0.05$ & $\bm{1.00 \pm 0.00}$ & $\bm{1.00 \pm 0.00}$ & $\bm{1.00 \pm 0.00}$ \\
    & $500\_\text{close}$ & $\bm{1.00 \pm 0.00}$ & $\bm{1.00 \pm 0.00}$ & $\bm{1.00 \pm 0.00}$ & $\bm{1.00 \pm 0.00}$ & $0.99 \pm 0.02$ & $0.99 \pm 0.02$ & $\bm{1.00 \pm 0.00}$ & $0.99 \pm 0.02$ \\
    \cmidrule(lr){2-10}
    & Average & $\bm{1.00 \pm 0.00}$ & $0.99 \pm 0.01$ & $\bm{1.00 \pm 0.00}$ & $\bm{1.00 \pm 0.00}$ & $0.98 \pm 0.03$ & $0.98 \pm 0.03$ & $0.98 \pm 0.04$ & $0.98 \pm 0.04$ \\

%    & \% of Optimal & 22.2 (2/9) & 33.3 (3/9) & 55.6 (5/9) & 66.7 (6/9) & 33.3 (3/9) & 33.3 (3/9) & 33.3 (3/9) & 44.4 (4/9) \\
    \midrule

\multirow{25}{*}{\rotatebox{90}{\emph{SIG SL}}} 
& \multicolumn{9}{l}{\textit{Without Fast Fading (NFF) $4\text{ag}$}} \\
\cmidrule(lr){2-10}
& $35\_\text{far}$    & $0.96 \pm 0.01$ & $0.97 \pm 0.00$ & $0.98 \pm 0.01$ & $0.98 \pm 0.01$ & $0.98 \pm 0.01$ & $\bm{0.99 \pm 0.01}$ & $0.98 \pm 0.03$ & $0.98 \pm 0.03$ \\
& $35\_\text{mid}$    & $0.95 \pm 0.02$ & $0.96 \pm 0.01$ & $0.98 \pm 0.01$ & $0.97 \pm 0.02$ & $0.98 \pm 0.01$ & $0.98 \pm 0.01$ & $\bm{1.00 \pm 0.01}$ & $0.98 \pm 0.06$ \\
& $35\_\text{close}$  & $0.98 \pm 0.01$ & $1.00 \pm 0.00$ & $1.01 \pm 0.00$ & $1.01 \pm 0.01$ & $1.02 \pm 0.01$ & $1.02 \pm 0.01$ & $1.02 \pm 0.01$ & $\bm{1.03 \pm 0.00}$ \\
& $123\_\text{far}$   & $0.99 \pm 0.00$ & $0.99 \pm 0.00$ & $0.99 \pm 0.00$ & $0.99 \pm 0.00$ & $0.99 \pm 0.00$ & $0.99 \pm 0.00$ & $\bm{1.00 \pm 0.00}$ & $\bm{1.00 \pm 0.00}$ \\
& $123\_\text{mid}$   & $0.97 \pm 0.03$ & $0.99 \pm 0.00$ & $1.01 \pm 0.01$ & $1.01 \pm 0.01$ & $1.00 \pm 0.02$ & $1.01 \pm 0.02$ & $\bm{1.03 \pm 0.00}$ & $\bm{1.03 \pm 0.01}$ \\
& $123\_\text{close}$ & $0.96 \pm 0.02$ & $0.99 \pm 0.01$ & $1.02 \pm 0.01$ & $1.02 \pm 0.01$ & $\bm{1.03 \pm 0.02}$ & $\bm{1.03 \pm 0.02}$ & $\bm{1.03 \pm 0.00}$ & $0.92 \pm 0.09$ \\
& $500\_\text{far}$   & $0.93 \pm 0.01$ & $0.94 \pm 0.00$ & $0.94 \pm 0.01$ & $0.94 \pm 0.01$ & $0.93 \pm 0.00$ & $0.93 \pm 0.00$ & $\bm{0.96 \pm 0.00}$ & $0.95 \pm 0.02$ \\
& $500\_\text{mid}$   & $0.99 \pm 0.00$ & $0.99 \pm 0.00$ & $0.99 \pm 0.00$ & $0.99 \pm 0.00$ & $\bm{1.00 \pm 0.00}$ & $\bm{1.00 \pm 0.00}$ & $\bm{1.00 \pm 0.00}$ & $\bm{1.00 \pm 0.01}$ \\
& $500\_\text{close}$ & $0.99 \pm 0.04$ & $1.08 \pm 0.05$ & $1.08 \pm 0.02$ & $1.02 \pm 0.07$ & $0.95 \pm 0.00$ & $0.95 \pm 0.00$ & $\bm{1.10 \pm 0.12}$ & $0.99 \pm 0.00$ \\
\cmidrule(lr){2-10}
& Average & $0.97 \pm 0.02$ & $0.99 \pm 0.04$ & $1.00 \pm 0.03$ & $0.99 \pm 0.02$ & $0.99 \pm 0.03$ & $0.99 \pm 0.03$ & $\bm{1.01 \pm 0.04}$ & $0.99 \pm 0.03$ \\

\cmidrule(lr){2-10}
& \multicolumn{9}{l}{\textit{With Fast Fading (FF) $4\text{ag}$}} \\
\cmidrule(lr){2-10}
& Average & $0.94 \pm 0.02$ & $0.94 \pm 0.02$ & $0.91 \pm 0.03$ & $0.91 \pm 0.03$ & $0.90 \pm 0.03$ & $0.91 \pm 0.03$ & $\bm{0.98 \pm 0.02}$ & $0.97 \pm 0.02$ \\

\cmidrule(lr){2-10}
& \multicolumn{9}{l}{\textit{With Fast Fading (FF) $8\text{ag}$}} \\
\cmidrule(lr){2-10}
& Average & $0.89 \pm 0.05$ & $0.88 \pm 0.04$ & $0.86 \pm 0.07$ & $0.83 \pm 0.08$ & $0.90 \pm 0.06$ & $0.90 \pm 0.06$ & $\bm{0.96 \pm 0.01}$ & $\bm{0.96 \pm 0.02}$ \\

\cmidrule(lr){2-10}
& \multicolumn{9}{l}{\textit{With Fast Fading (FF) $16\text{ag}$}} \\
\cmidrule(lr){2-10}
& Average & $0.82 \pm 0.09$ & $0.82 \pm 0.11$ & $0.88 \pm 0.07$ & $0.83 \pm 0.08$ & $0.90 \pm 0.05$ & $0.89 \pm 0.05$ & $\bm{0.95 \pm 0.03}$ & $\bm{0.95 \pm 0.03}$ \\

\midrule
    
\multirow{3}{*}{\rot{\emph{SIG ML}}} 
    & $4\text{ag}$  & $0.84 \pm 0.03$ & $0.85 \pm 0.03$ & $0.88 \pm 0.04$ & $0.86 \pm 0.01$ & $0.92 \pm 0.02$ & $0.92 \pm 0.03$ & $\bm{0.95 \pm 0.01}$ & $\bm{0.95 \pm 0.01}$ \\
    & $8\text{ag}$  & $0.58 \pm 0.05$ & $0.56 \pm 0.01$ & $0.70 \pm 0.05$ & $0.58 \pm 0.04$ & $0.84 \pm 0.02$ & $0.84 \pm 0.04$ & $\bm{0.92 \pm 0.01}$ & $\bm{0.92 \pm 0.01}$ \\
    & $16\text{ag}$ & $0.35 \pm 0.01$ & $0.33 \pm 0.03$ & $0.35 \pm 0.02$ & $0.38 \pm 0.01$ & $0.58 \pm 0.01$ & $0.59 \pm 0.04$ & $\bm{0.85 \pm 0.02}$ & $0.84 \pm 0.01$ \\
    \midrule
    
\multirow{3}{*}{\rot{\emph{POSIG}}}
  & $4\text{ag}$  & $0.80 \pm 0.02$ & $0.81 \pm 0.03$ & $0.82 \pm 0.02$ & $0.81 \pm 0.02$ & $0.84 \pm 0.03$ & $0.88 \pm 0.04$ & $0.89 \pm 0.02$ & $\bm{0.90 \pm 0.02}$ \\
  & $8\text{ag}$  & $0.64 \pm 0.03$ & $0.64 \pm 0.01$ & $0.70 \pm 0.03$ & $0.69 \pm 0.01$ & $0.81 \pm 0.02$ & $0.85 \pm 0.03$ & $0.86 \pm 0.01$ & $\bm{0.88 \pm 0.03}$ \\
  & $16\text{ag}$ & $0.55 \pm 0.03$ & $0.52 \pm 0.01$ & $0.60 \pm 0.02$ & $0.53 \pm 0.02$ & $0.79 \pm 0.02$ & $0.84 \pm 0.01$ & $0.84 \pm 0.01$ & $\bm{0.87 \pm 0.01}$ \\
  
    \bottomrule
  \end{tabular}
  \vspace{0.3em}

\label{tab:benchmarking}
\end{table*}

\section{Results and Discussions}  

\subsection{NFIG}
Table~\ref{tab:tasks_summary} shows that all eight algorithms achieve near-optimal performance in \emph{NFIG}, with average normalized returns of $0.98$--$1.00$. A sensitivity analysis using alternative reward weightings of $\lambda_1{:}\lambda_2=0.5{:}0.5$ and $0.25{:}0.75$ yields average normalized returns above $0.95$ for all eight algorithms, confirming that this near-optimal performance is not specific to the adopted weighting.

The results indicate that non-stationarity has only a limited impact in the evaluated \emph{NFIG} tasks. Despite simultaneous learning among agents, all value-based and actor-critic methods achieve near-optimal performance, with the averages of the former larger than the latter by at most $0.02$. This suggests that, in the evaluated four-agent, fixed-topology, one-step setting, non-stationarity does not substantially differentiate the performance of the learning mechanisms discussed in Section~V.C.

Coordination difficulty is also relatively mild. Since all agents share the same reward, a globally optimal joint action is a Nash equilibrium (NE), but suboptimal NEs may also exist and cannot necessarily be escaped through unilateral action changes. Our enumeration shows that 96\% of the learned joint actions are pure NEs. Although substantially suboptimal NEs exist, with normalized returns as low as $0.73$, the algorithms generally avoid these poor equilibria. The suboptimal NEs they reach are typically near-optimal, with an average normalized-return gap of only $0.03$.

Value decomposition helps coordination in \emph{NFIG}. VDN and QMIX reach the optimal joint action in 91\% of the runs, compared with 68\% for IDQN and Hys-IDQN. Because the suboptimal NEs reached by the independent learners are close to the optimum, this advantage yields only a small gain in return. In contrast, MAA2C and MAPPO offer no advantage over their IL counterparts in \emph{NFIG}. As discussed in Section~V.C, centralized critics provide little additional information in fully observable tasks, while decentralized actors must still learn to coordinate their actions.

Actor-critic methods perform slightly worse than value-based methods, mainly because they reach the optimal joint action in only about half of the runs. Most of the remaining runs converge to suboptimal pure NEs, which are generally near-optimal, so the performance differences remain small.

\textbf{The key insights from \emph{NFIG} results are:}
\begin{itemize}
\item \textbf{Coordination difficulty and non-stationarity
have limited impact on learning performance in simple
settings with small network scale, fixed topology,
and one-step decision horizons, where all algorithms
generally converge to optimal or near-optimal solutions.}
\item \textbf{Value-based methods slightly outperform actor-critic methods. Value decomposition provides a small advantage over value-based IL, whereas centralized critics offer no advantage over their IL counterparts.}
\end{itemize}

\begin{figure*}[tb!]
	\centering 
\includegraphics[width=\textwidth]{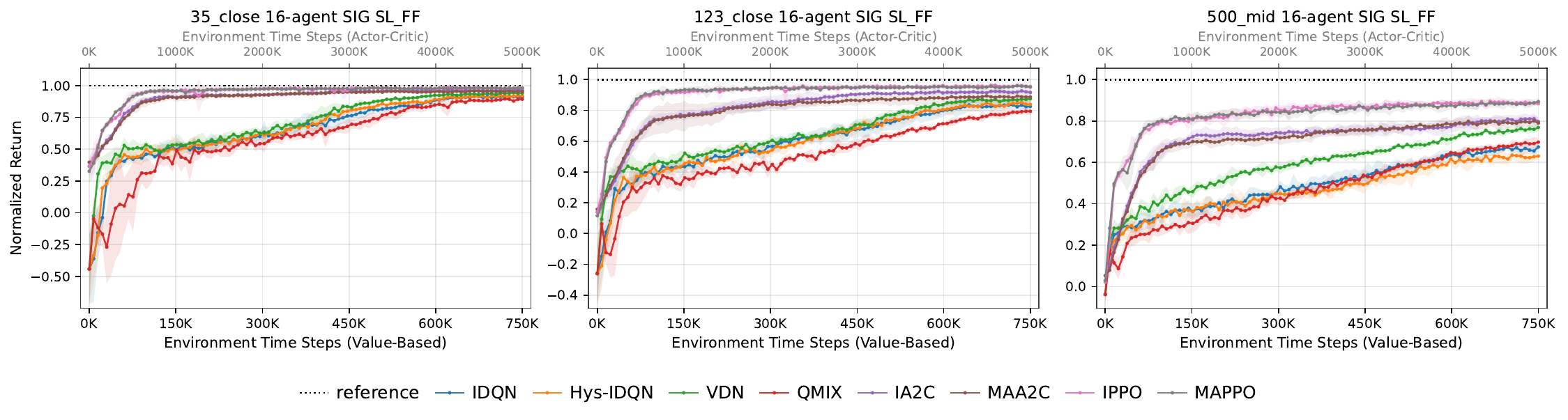}

	\caption{Normalized returns in selected 16-agent \emph{SIG SL\_FF} tasks. Shaded areas represent the 95\% CI over five seeds.}
	\label{fig:SIGSL}
\end{figure*}

\subsection{SIG SL}
Moving from the single-step \emph{NFIG} to the 50-step \emph{SIG SL\_NFF} has little impact on performance, with all eight algorithms achieving average normalized returns of at least $0.97$ (Table~\ref{tab:tasks_summary}). However, actor-critic methods generally perform comparably to or slightly better than value-based methods, reversing the trend in \emph{NFIG}. This is consistent with the analysis in Section~V.C, as multi-step returns may provide an advantage over short-horizon bootstrapped TD targets in sequential decision-making tasks.

Introducing fast fading causes moderate performance degradation in the four-agent \emph{SIG SL} tasks, reducing average normalized returns by $0.02$--$0.09$ across algorithms. Nevertheless, all algorithms maintain returns of at least $0.90$. PPO-based methods exhibit the smallest degradation, followed by value-based IL, whereas value-based CTDE and A2C-based methods are most affected. Fast fading introduces time-varying channel conditions at each time step, and PPO's relative robustness is consistent with the stabilizing effect of its clipped policy updates, as discussed in Section~V.C. The greater sensitivity of value-based CTDE and A2C warrants further investigation.

Increasing the number of agents in \emph{SIG SL\_FF} causes moderate performance degradation, with actor-critic methods increasingly outperforming value-based methods. At 16 agents, value-based methods achieve normalized returns of $0.82$--$0.88$, compared with $0.89$--$0.90$ for A2C-based methods and approximately $0.95$ for PPO-based methods. Increasing network scale expands the joint action space and intensifies non-stationarity and coordination difficulty. The growing advantage of actor-critic methods is consistent with the differences in training strategies discussed in Section~V.C, particularly their use of on-policy learning and PPO's clipped policy updates.

The effect of value decomposition also changes with network scale. Under fast fading, value decomposition performs worse than value-based IL at four and eight agents, but achieves higher average returns at 16 agents. This suggests that the relative benefit of value decomposition may increase as coordination becomes more challenging. As in \emph{NFIG}, centralized critics offer no advantage over their IL counterparts, consistent with the analysis in Section~V.C.

\textbf{The key insights from \emph{SIG SL} results are:}
\begin{itemize}
\item \textbf{Sequential decision making alone has little impact on learning performance, while fast fading and increasing network scale cause moderate performance degradation.}

\item \textbf{PPO-based methods are most robust to fast fading and increasing network scale, while actor-critic methods increasingly outperform value-based methods as network scale grows. Under fast fading, value decomposition underperforms value-based IL at four and eight agents but achieves better performance at 16 agents, while centralized critics provide no advantage over IL.}
\end{itemize}

\begin{figure*}[tb!]
	\centering 
\includegraphics[width=\textwidth]{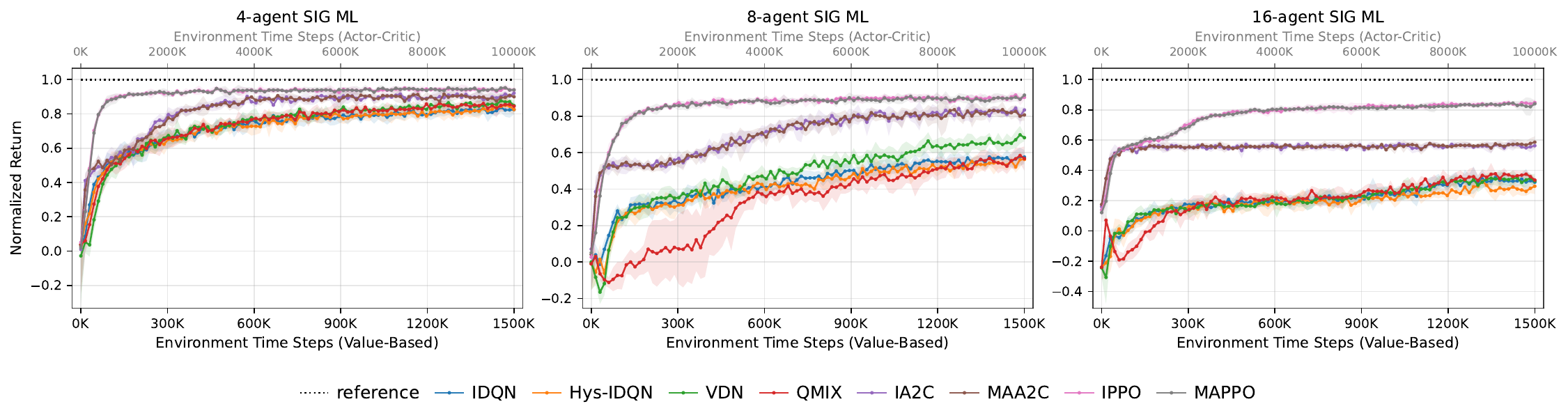}

	\caption{Normalized returns in \emph{SIG ML} tasks. Shaded areas represent the 95\% CI over five seeds.}
	\label{fig:SIGML}
\end{figure*}

\subsection{SIG ML}

\emph{SIG ML} trains a single policy across 15,000 topologies in the 4-agent setting and 60,000 in the 8- and 16-agent settings, and evaluates it on the nine representative topologies used in \emph{SIG SL}. Topology variation introduces substantial additional learning difficulty as the number of agents increases. Relative to \emph{SIG SL\_FF}, the normalized returns of all algorithms change by at most $0.10$ in the 4-agent setting, whereas at 16 agents value-based methods fall from $0.82$--$0.88$ to $0.33$--$0.38$, and even PPO-based methods fall from $0.95$ to $0.84$--$0.85$ (Table~\ref{tab:tasks_summary}). Since \emph{SIG SL\_FF} has the same number of agents and joint action space, this degradation is specifically associated with learning across varying topologies.\par

Two challenges could drive this degradation: robustness, i.e., learning a single policy that performs well across diverse training topologies, and generalization, i.e., transferring that policy to topologies unseen during training. To separate these effects, we conduct an additional controlled experiment for IDQN and IPPO in the 16-agent setting. We randomly sample one seen topology from the \emph{SIG ML} training set for each combination of three vehicle-density levels (35, 123, and 500~veh/km) and three vehicle-to-BS distance levels (\texttt{close}, \texttt{mid}, and \texttt{far}), forming a $3\times3$ matrix matched to the nine evaluation topologies. We then train a separate \emph{SIG SL} policy for each seen topology.

As shown in Table~\ref{tab:robustness_generalization}, we measure the robustness gap by comparing \emph{SIG SL} and \emph{SIG ML} on the same nine seen topologies ($A-C$), and the generalization gap by evaluating the same \emph{SIG ML} policy on the nine seen and nine evaluation topologies ($C-D$). The robustness gaps are $0.45$ and $0.09$ for IDQN and IPPO, respectively, whereas the generalization gaps are negligible for both algorithms. Moreover, \emph{SIG SL} achieves similar performance on the nine seen and nine evaluation topologies ($A\approx B$), linking this controlled analysis to the \emph{SIG SL\_FF}--\emph{SIG ML} degradation observed in Table~\ref{tab:tasks_summary}. These results indicate that the degradation is mainly associated with the difficulty of learning a single robust policy across diverse topologies, rather than generalizing to unseen topologies under the same topology distributions.

\begin{table}[t]

\centering
\caption{Robustness and generalization analysis in the 16-agent \emph{SIG} setting.}
\label{tab:robustness_generalization}
\begin{tabular}{@{}lcc@{}}
\toprule
 & IDQN & IPPO \\
\midrule
\emph{SIG SL}, 9 seen ($A$)       & 0.80 & 0.94 \\
\emph{SIG ML}, 9 seen ($C$)       & 0.35 & 0.85 \\
\emph{SIG SL}, 9 evaluation ($B$) & 0.82 & 0.95 \\
\emph{SIG ML}, 9 evaluation ($D$) & 0.35 & 0.85 \\
\midrule
Robustness gap ($A-C$)             & 0.45 & 0.09 \\
Generalization gap ($C-D$)         & 0.00 & 0.00 \\
\bottomrule
\end{tabular}
\end{table}

Actor-critic methods handle topology variation substantially better than value-based methods, and the gap widens with network scale. The best actor-critic method exceeds the best value-based method by $0.07$ in the 4-agent setting and by $0.47$ in the 16-agent setting. Within the actor-critic family, the advantage of PPO-based over A2C-based methods also increases with the number of agents, from $0.03$ at 4 agents to $0.08$ at 8 agents and $0.26$ at 16 agents. Centralized training again provides limited benefit. MAA2C and MAPPO remain within $0.01$ of their IL counterparts at every scale, consistent with the fully observable setting leaving centralized critics little additional information. Value decomposition generally provides modest gains over independent value-based learning.

\textbf{The key insights from \emph{SIG ML} results are:}
\begin{itemize}
\item \textbf{Learning robust policies across diverse vehicular topologies poses a major challenge at larger network scales. In contrast, generalization from seen to unseen topologies drawn from the same distribution introduces negligible additional performance degradation.}
\item \textbf{Actor-critic methods substantially outperform value-based methods in handling topology variation, with the performance gap widening as network scale increases. Value decomposition generally provides a modest advantage over value-based IL, whereas centralized critics offer no advantage over actor-critic IL in the fully observable \emph{SIG ML} tasks.}
\end{itemize}

\begin{figure*}[tb!]
	\centering 
\includegraphics[width=\textwidth]{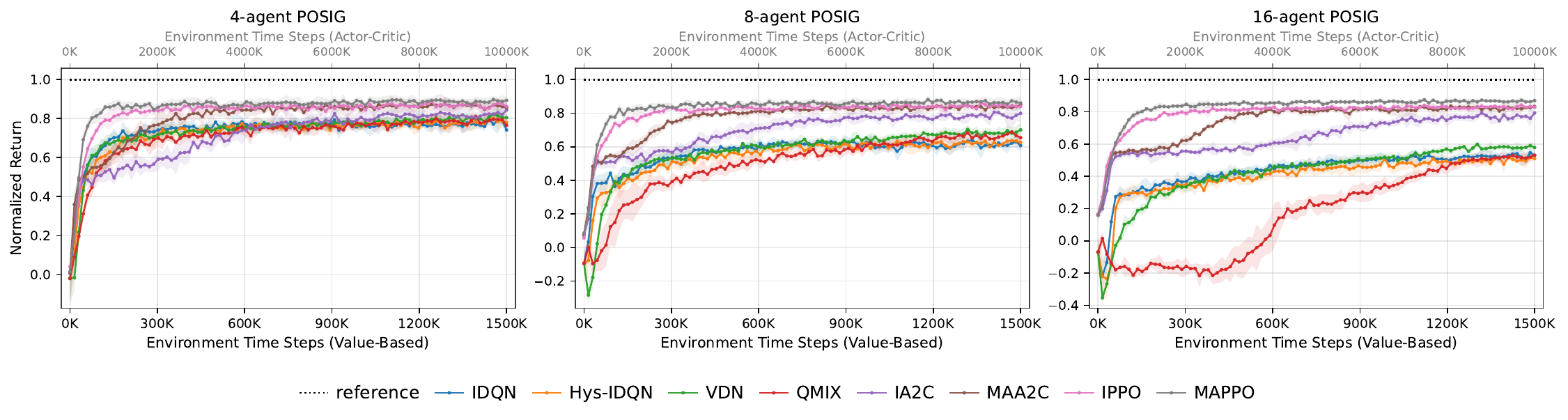}

	\caption{Normalized returns in \emph{POSIG} tasks. Shaded areas represent the 95\% CI over five seeds.}
	\label{fig:POSIG}
\end{figure*}

\subsection{POSIG}

The performance difference between \emph{SIG ML} and \emph{POSIG} depends strongly on network scale (Table~\ref{tab:tasks_summary}). In the 4-agent setting, using local observations instead of the global state reduces the normalized return of every algorithm by $0.04$--$0.08$. In the 8-agent setting, the changes range from $-0.06$ to $+0.11$ without a consistent direction, whereas in the 16-agent setting all algorithms except IPPO ($-0.01$) achieve higher returns in \emph{POSIG}, with gains of up to $0.25$ for VDN and MAA2C. These results suggest a scale-dependent trade-off in the use of global information. While non-local information can facilitate coordination, exploiting it may become increasingly difficult as network scale and topology diversity increase. Consequently, restricting agents to local information may simplify learning across diverse topologies despite the resulting partial observability.

However, the \emph{SIG ML}--\emph{POSIG} comparison also changes the input dimension, so the improvement at larger scales cannot be attributed to the exclusion of non-local information alone. To separate the two factors, we set the state components outside each agent's local observation to zero in the 16-agent \emph{SIG ML} setting for IDQN, VDN, IPPO, and MAPPO. This masked global state provides the same information as the local observation while preserving the global-state input dimension, network architecture, and training configuration. As shown in Table~\ref{tab:masked_state}, masking the non-local components increases the normalized return by $0.20$ and $0.25$ for IDQN and VDN and by $0.02$ and $0.05$ for IPPO and MAPPO, whereas further reducing the input to the local observation changes it by at most $0.01$.

Therefore, the improvement in the 16-agent setting is mainly associated with excluding non-local information rather than with the reduction in input dimension. 

\begin{table}[t]

\centering
\caption{Normalized returns with different input representations in the 16-agent setting. IDQN and VDN are trained for $1.5$ million environment time steps, and IPPO and MAPPO for $5$ million time steps.}
\label{tab:masked_state}
\footnotesize
\setlength{\tabcolsep}{6pt}
\renewcommand{\arraystretch}{1.1}
\begin{tabular}{@{}lcccc@{}}
\toprule
\textbf{Input} & \textbf{IDQN} & \textbf{VDN} & \textbf{IPPO} & \textbf{MAPPO} \\
\midrule
Global state (\emph{SIG ML})       & 0.35 & 0.35 & 0.81 & 0.82 \\
Masked global state                & 0.55 & 0.60 & 0.83 & 0.87 \\
Local observation (\emph{POSIG})   & 0.55 & 0.60 & 0.84 & 0.87 \\
\bottomrule
\end{tabular}
\end{table}

Actor-critic methods outperform value-based methods in \emph{POSIG} at every network scale, and the gap widens with the number of agents. With four agents, value-based and actor-critic methods achieve normalized returns of $0.80$--$0.82$ and $0.84$--$0.90$, respectively, whereas with 16 agents value-based methods fall to $0.52$--$0.60$ and actor-critic methods remain at $0.79$--$0.87$. MAPPO achieves the highest return in all three settings, followed closely by IPPO.

Unlike in the fully observable tasks, centralized-critic methods outperform their IL counterparts under partial observability, with MAA2C exceeding IA2C by $0.04$--$0.05$ and MAPPO exceeding IPPO by $0.01$--$0.03$. This is consistent with the analysis in Section~V.C, as the centralized critic can exploit global information unavailable to an independent critic. Value decomposition also provides a modest advantage over value-based IL.

\textbf{The key insights from the \emph{POSIG} results are:}
\begin{itemize}
\item \textbf{The effect of restricting agents to local information depends strongly on network scale. Local observations reduce performance in the 4-agent setting but outperform the global-state setting for most algorithms with 16 agents. }

\item \textbf{Actor-critic methods, particularly PPO-based methods, remain the most effective under partial observability. Unlike in the fully observable tasks, centralized critics consistently outperform their IL counterparts, although the advantage for PPO remains small, whereas value decomposition provides a modest advantage.}
\end{itemize}

The learning curves for the 16-agent \emph{SIG SL\_FF} task and for all \emph{SIG ML} and \emph{POSIG} tasks are shown in Figs.~\ref{fig:SIGSL}, \ref{fig:SIGML}, and \ref{fig:POSIG}, respectively. Actor-critic methods, particularly the PPO-based methods, learn more stably than value-based methods, with smoother learning curves and smaller variation across seeds. This is consistent with Section~V.C, as value estimation errors can propagate through the bootstrapped TD targets of value-based methods, whereas PPO constrains policy changes through its clipped surrogate objective. The PPO-based methods also exceed the highest returns of the value-based methods early in training, and the performance gaps between the algorithm classes persist until the end of training. Moreover, over the last 10\% of training, the average normalized return of every algorithm changes by at most $0.05$, indicating that the selected training horizons are generally sufficient for the algorithms to approach stable performance.

\subsection{Summary of Cross-Task and Cross-Algorithm Analysis}

To summarize the findings of Section~VII,
Table~\ref{tab:attribution} quantifies the incremental performance changes associated with the MARL challenges introduced across the task hierarchy.
Each entry reports the change in average normalized
return across the eight algorithms, in percentage
points (pp), relative to the corresponding baseline
task. Since these challenges may interact, the
performance changes reflect their combined effects
rather than the independent contribution of each
additional challenge.

For \emph{NFIG}, performance is compared with the
optimal solution. The transition to \emph{SIG~SL\_NFF}
introduces sequential decision making along with a
different reward formulation. Environmental stochasticity
is evaluated by comparing the 4-agent
\emph{SIG~SL\_FF} and \emph{SIG~SL\_NFF} tasks.
The effect of a larger joint action space is examined
by comparing the 8- and 16-agent \emph{SIG~SL\_FF}
tasks with their 4-agent counterpart. These cross-scale
comparisons use scale-specific normalization references.
The remaining rows compare consecutive tasks with
the same number of agents.

\begin{table}[t]
\centering
\caption{Incremental normalized-return changes
associated with progressively introduced MARL challenges.}
\label{tab:attribution}
\footnotesize
\setlength{\tabcolsep}{3pt}
\renewcommand{\arraystretch}{1.2}
\begin{tabular}{@{}llccc@{}}
\toprule
\textbf{Task} & \makecell[l]{\textbf{Principal additional}\\\textbf{MARL challenge}}
& \multicolumn{3}{c}{\textbf{Performance change (pp)}} \\
\cmidrule(l){3-5}
& & \textbf{4\,ag.} & \textbf{8\,ag.}
& \textbf{16\,ag.} \\
\midrule
\emph{NFIG}
& Coordination, non-stationarity
& $-1$ & --- & --- \\
\emph{SIG~SL\_NFF}
& Sequential decision making
& $0$ & --- & --- \\
\emph{SIG~SL\_FF}
& Environmental stochasticity
& $-6$ & --- & --- \\
\emph{SIG~SL\_FF}
& Larger joint action space
& --- & $-4$ & $-5$ \\
\emph{SIG~ML}
& Robustness, generalization
& $-4$ & $-16$ & $-35$ \\
\emph{POSIG}
& Partial observability
& $-5$ & $+2$ & $+16$ \\
\bottomrule
\end{tabular}
\end{table}

The results show that coordination, non-stationarity,
and sequential decision making do not prevent the
algorithms from achieving high normalized returns
in the simplest tasks. In contrast, introducing
environmental stochasticity and enlarging the joint
action space are associated with more noticeable
performance degradation. The impact associated with
robustness and generalization challenges increases
substantially with network scale: moving to
\emph{SIG ML} reduces the average normalized return
by $4$, $16$, and $35$~pp for 4, 8, and 16 agents,
respectively. Together with the controlled
robustness--generalization analysis, these results
identify learning robust policies across diverse
topologies as a major performance bottleneck at
larger scales, while the additional generalization
gap to unseen topologies drawn from the same
distribution is negligible in the evaluated setting.

Switching to local observations reduces average performance by $5$~pp at 4 agents, but improves it by $2$~pp at 8 agents and $16$~pp at 16 agents.
This does not imply that partial observability is
intrinsically beneficial at larger scales. Instead,
the results suggest that excluding non-local information
may simplify policy learning across diverse topologies
enough to offset the associated information loss.

Table~\ref{tab:algorithm_comparison} further compares the MARL paradigms across the task hierarchy. AC--VB compares the average performance of the four actor-critic methods with the four value-based methods; VD--IL compares VDN/QMIX with IDQN/Hys-IDQN; and CC--IL compares MAA2C/MAPPO with IA2C/IPPO. Positive values indicate an advantage of the first paradigm.

\begin{table}[t]

\centering
\caption{Cross-algorithm comparison of normalized-return performance.}
\label{tab:algorithm_comparison}
\footnotesize
\setlength{\tabcolsep}{5pt}
\renewcommand{\arraystretch}{1.15}
\begin{tabular}{@{}lcrrr@{}}
\toprule
\textbf{Task} & \textbf{Agents}
& \multicolumn{1}{c}{\textbf{AC--VB}}
& \multicolumn{1}{c}{\textbf{VD--IL}}
& \multicolumn{1}{c}{\textbf{CC--IL}} \\
& & \multicolumn{3}{c}{\textbf{Performance advantage (pp)}} \\
\midrule
\emph{NFIG}        & 4  & $-1.8$ & $+0.5$ & $0.0$ \\
\emph{SIG SL\_NFF} & 4  & $+0.8$ & $+1.5$ & $-1.0$ \\
\emph{SIG SL\_FF}  & 4  & $+1.5$ & $-3.0$ & $0.0$ \\
                    & 8  & $+6.5$ & $-4.0$ & $0.0$ \\
                    & 16 & $+8.5$ & $+3.5$ & $-0.5$ \\
\emph{SIG ML}      & 4  & $+7.8$ & $+2.5$ & $0.0$ \\
                    & 8  & $+27.5$ & $+7.0$ & $0.0$ \\
                    & 16 & $+36.3$ & $+2.5$ & $0.0$ \\
\emph{POSIG}       & 4  & $+6.8$ & $+1.0$ & $+2.5$ \\
                    & 8  & $+18.3$ & $+5.5$ & $+3.0$ \\
                    & 16 & $+28.5$ & $+3.0$ & $+4.0$ \\
\bottomrule
\end{tabular}
\end{table}

The cross-algorithm comparison reveals two main trends. First, the advantage of actor-critic over value-based learning grows markedly with network scale and is most pronounced in the tasks with topology variation, reaching $36.3$~pp in the 16-agent \emph{SIG ML} task and $28.5$~pp in the 16-agent \emph{POSIG} task. Second, centralized critics provide no advantage over their IL counterparts in the fully observable tasks but outperform them by $2.5$--$4.0$~pp in \emph{POSIG}, whereas value decomposition provides a modest advantage over value-based IL in both \emph{SIG ML} and \emph{POSIG}, by $1.0$--$7.0$~pp. Most of these empirical findings are consistent with the algorithmic analysis in Section~V.C, while the greater sensitivity of value decomposition to fast fading, the substantial performance degradation associated with learning across diverse topologies at larger network scales, and the scale-dependent benefits of excluding non-local information warrant further investigation.

\section{Conclusion and Future Works}

In this paper, we proposed a hierarchy of multi-agent interference games for C-V2X RRA that progressively introduces key MARL challenges. Using the resulting benchmark framework, we evaluated eight representative MARL algorithms and obtained the following insights:

\begin{itemize}
\item Among the challenges considered, learning robust policies across diverse vehicular topologies poses the greatest difficulty at larger network scales.

\item The relative advantage of actor-critic methods over value-based methods widens as the network scale increases, most notably when the policy must cope with topology variation.

\item In the most challenging \emph{POSIG} tasks, PPO-based methods achieve the strongest overall performance, with MAPPO attaining the highest returns and IPPO following closely. Considering the scalability requirements of centralized critics, IPPO offers a compelling balance between learning performance and implementation complexity for large-scale C-V2X RRA.
\end{itemize}

While most observed differences among MARL paradigms are consistent with the algorithmic analysis in Section~V.C, the mechanisms underlying their varying robustness to vehicular topology diversity remain insufficiently understood. Understanding these mechanisms and developing more robust MARL algorithms are important directions for future research.

The scale-dependent \emph{SIG ML}--\emph{POSIG} results further highlight the importance of information and state representation. At larger scales, the benefits of simplifying policy learning by excluding non-local information may outweigh the associated information loss. This motivates further investigation of representation-learning approaches, such as GNN-based MARL \cite{ji2025graphneuralnetworksdeep} and transformer-based MARL, as well as other mechanisms for selectively exploiting task-relevant information. The proposed task hierarchy provides a controlled platform for determining which MARL challenges these approaches address, rather than evaluating only their aggregate performance in complex environments.

Several extensions can further broaden the scope of the benchmark. The current experiments consider up to 16 simultaneously coordinating agents and adopt controlled assumptions including noise-free observations, instantaneous channel state information, and a fixed set of active agents. Future studies can extend the benchmark to larger networks and introduce additional sources of complexity, such as noisy or delayed observations and missing or inactive agents, as separate task variants while keeping other factors controlled. \par

Finally, the results are obtained under the standardized 3GPP TR~36.885 highway evaluation framework, with topology variation under a fixed channel model. Changes in propagation environments, channel models, mobility patterns, and traffic characteristics may introduce additional distribution shifts. Extending the benchmark to these settings would help assess whether topology robustness remains the dominant challenge and whether the relative performance of MARL paradigms generalizes across deployment conditions. \par

\begin{comment}
{\color{blue}\sout{Furthermore, the proposed benchmark framework is not tied to the considered highway setting. It can be readily applied to other C-V2X deployment scenarios by replacing the topology dataset and channel model while preserving the underlying \mbox{\emph{NFIG}}, \mbox{\emph{SIG}}, and \mbox{\emph{POSIG}} task structure. This provides a systematic pathway for extending the benchmark to urban and other deployment scenarios in future work, as well as for evaluating the relative impact of topology-induced and channel-model-induced distribution shifts on MARL performance. Such studies would help establish the extent to which the findings reported in this paper generalize across different C-V2X environments.}}
\end{comment}

%The impact of a larger action space, longer time horizons, sparse rewards, and more diverse interference scenarios requires further study in the future. \par

%Although some efforts have been made to learn more efficient state representations using GNNs \cite{ji2025graphneuralnetworksdeep} and to improve generalization through meta-learning \cite{Yuan2021MetaRLV2X}, further research is still needed. 

%, the current benchmark provides valuable insights into the relative importance of different MARL challenges and their implications for future algorithm design in C-V2X resource allocation.
\begin{appendices}

\section{Hyperparameters}
The hyperparameters used in all experiments are given in Tables~\ref{tab:hyperparameters_vb_nfig_sigsl}--\ref{tab:hyperparameters_ac_sigml_posig}.
\begin{table*}[!htbp]
  \centering
  \caption{\emph{NFIG} / \emph{SIG SL} Hyperparameters for Value-Based Algorithms}
  \label{tab:hyperparameters_vb_nfig_sigsl}
  \small
  \begin{tabular}{@{}p{4.5cm}>{\centering\arraybackslash}p{2.5cm}>{\centering\arraybackslash}p{2.5cm}>{\centering\arraybackslash}p{2.5cm}>{\centering\arraybackslash}p{2.5cm}@{}}
    \toprule
    \textbf{Hyperparameter} & \textbf{IDQN} & \textbf{Hys-IDQN} & \textbf{VDN} & \textbf{QMIX} \\
    \midrule
    hidden dimension (FC)        & $128$           & $128$           & $128$           & $128$ \\
    learning rate            & $3 \times 10^{-5}$          & $3 \times 10^{-5}$          & $3 \times 10^{-5}$          & $3 \times 10^{-5}$ \\
    mixing network learning rate & -       & -           & -           & $1 \times 10^{-6}$ \\
    hysteretic learning rate & -           & $\alpha{=}1.0, \beta{=}0.2$ & - & - \\
    batch size               & $64$         & $64$         & $64$         & $64$ \\
    target update (soft)     & $5 \times 10^{-3}$          & $5 \times 10^{-3}$          & $5 \times 10^{-3}$          & $5 \times 10^{-3}$ \\
    training time steps (million) & $0.05$ / $0.75$ & $0.05$ / $0.75$ & $0.05$ / $0.75$ & $0.05$ / $0.75$ \\
    epsilon anneal time steps (million, linear) & $0.04$ / $0.6$ & $0.04$ / $0.6$ & $0.04$ / $0.6$ & $0.04$ / $0.6$ \\
    parameter sharing        & False         & False         & False         & False \\
    \bottomrule
    \multicolumn{5}{p{14.5cm}}{\footnotesize\textit{Note:} Values shown as $a / b$ indicate different settings for \emph{NFIG} / \emph{SIG SL} respectively. Single values apply to both. For QMIX in the 16-agent \emph{SIG SL} task, the mixing network learning rate is $1 \times 10^{-5}$.} \\
  \end{tabular}
\end{table*}

\begin{table*}[!htbp]
  \centering
  \caption{\emph{NFIG} / \emph{SIG SL} Hyperparameters for Actor-Critic Algorithms}
  \label{tab:hyperparameters_ac_nfig_sigsl}
  \small
  \begin{tabular}{@{}p{4.5cm}>{\centering\arraybackslash}p{2.5cm}>{\centering\arraybackslash}p{2.5cm}>{\centering\arraybackslash}p{2.5cm}>{\centering\arraybackslash}p{2.5cm}@{}}
    \toprule
    \textbf{Hyperparameter} & \textbf{IA2C} & \textbf{MAA2C} & \textbf{IPPO} & \textbf{MAPPO} \\
    \midrule
    hidden dimension (FC)         & $128$           & $128$           & $128$           & $128$ \\
    actor learning rate ($\alpha$)     & $2 \times 10^{-4}$          & $2 \times 10^{-4}$          & $4 \times 10^{-4}$          & $4 \times 10^{-4}$ \\
    critic learning rate ($\beta$)     & $2 \times 10^{-4}$          & $2 \times 10^{-4}$          & $6 \times 10^{-4}$          & $6 \times 10^{-4}$ \\
    batch size               & $8$             & $8$             & $256$ / $64$    & $256$ / $64$ \\
    mini-batches             & -           & -           & $4$             & $4$ \\
    target update (soft)     & $0.01$          & $0.01$          & -           & - \\
    PPO epochs               & -           & -           & $10$            & $10$ \\
    entropy coefficient      & -           & -           & $0.01$ / $0.001$ & $0.01$ / $0.001$ \\
    training time steps (million) & $0.05$ / $5$ & $0.05$ / $5$ & $0.05$ / $5$ & $0.05$ / $5$ \\
    parameter sharing        & False         & False         & True         & True \\
    \bottomrule
\multicolumn{5}{p{14.5cm}}{\footnotesize\textit{Note:} Values shown as $a / b$ indicate different settings for \emph{NFIG} / \emph{SIG SL} respectively. Single values apply to both.} \\
  \end{tabular}
\end{table*}
\begin{table*}[!htbp]
  \centering
  \caption{\emph{SIG ML} / \emph{POSIG} Hyperparameters for Value-Based Algorithms}
  \label{tab:hyperparameters_vb_sigml_posig}
  \small
  \begin{tabular}{@{}p{4.5cm}>{\centering\arraybackslash}p{2.5cm}>{\centering\arraybackslash}p{2.5cm}>{\centering\arraybackslash}p{2.5cm}>{\centering\arraybackslash}p{2.5cm}@{}}
    \toprule
    \textbf{Hyperparameter} & \textbf{IDQN} & \textbf{Hys-IDQN} & \textbf{VDN} & \textbf{QMIX} \\
    \midrule
    hidden dimension (FC)         & $128$           & $128$           & $128$           & $128$ \\
    learning rate            & $1 \times 10^{-5}$      & $1 \times 10^{-5}$      & $1 \times 10^{-5}$          & $1 \times 10^{-5}$ \\
    mixing network learning rate & -       & -           & -           & $1 \times 10^{-6}$ \\
    hysteretic learning rate & -           & $\alpha{=}1.0, \beta{=}0.2$ & - & - \\
    batch size               & $64$         & $64$         & $64$         & $64$ \\
    target update (soft)     & $5 \times 10^{-3}$          & $5 \times 10^{-3}$          & $5 \times 10^{-3}$          & $5 \times 10^{-3}$ \\
    training time steps (million) & $1.5$ & $1.5$ & $1.5$ & $1.5$ \\
    epsilon anneal time steps (million, linear) & $1.2$ & $1.2$ & $1.2$ & $1.2$ \\
    parameter sharing        & False        & False         & False         & False \\
    \bottomrule
    \multicolumn{5}{p{14.5cm}}{\footnotesize\textit{Note:} For VDN in \emph{SIG ML}, the learning rate is $5 \times 10^{-6}$.} \\
  \end{tabular}
\end{table*}

\begin{table*}[!htbp]
  \centering
  \caption{\emph{SIG ML} / \emph{POSIG} Hyperparameters for Actor-Critic Algorithms}
  \label{tab:hyperparameters_ac_sigml_posig}
  \small
  \begin{tabular}{@{}p{4.5cm}>{\centering\arraybackslash}p{2.5cm}>{\centering\arraybackslash}p{2.5cm}>{\centering\arraybackslash}p{2.5cm}>{\centering\arraybackslash}p{2.5cm}@{}}
    \toprule
    \textbf{Hyperparameter} & \textbf{IA2C} & \textbf{MAA2C} & \textbf{IPPO} & \textbf{MAPPO} \\
    \midrule
    hidden dimension (FC)         & $128$           & $128$           & $128$           & $128$ \\
    actor learning rate ($\alpha$)     & $5 \times 10^{-4}$          & $5 \times 10^{-4}$          & $4 \times 10^{-4}$          & $4 \times 10^{-4}$ \\
    critic learning rate ($\beta$)     & $5 \times 10^{-4}$          & $5 \times 10^{-4}$          & $6 \times 10^{-4}$          & $6 \times 10^{-4}$ \\
    batch size               & $8$             & $8$             & $256$           & $256$ \\
    mini-batches             & -           & -           & $4$             & $4$ \\
    target update (soft)     & $0.01$          & $0.01$          & -           & - \\
    PPO epochs               & -           & -           & $10$            & $10$ \\
    entropy coefficient      & -           & -           & $0.001$         & $0.001$ \\
    training time steps (million) & $10$ & $10$ & $10$ & $10$ \\
    parameter sharing        & True         & True         & True         & True \\
    \bottomrule
  \end{tabular}
\end{table*}

\end{appendices}
	
\bibliographystyle{IEEEtran}
\bibliography{reference}
	
\end{document}